 \documentclass[final,3p,review]{elsarticle}
\usepackage{amssymb}
\usepackage{amsmath}
\usepackage{lineno}
\usepackage[figuresright]{rotating}
\usepackage{graphicx}% Include figure files
\usepackage[figurename=Fig.]{caption}
\usepackage{float}
\usepackage{tabu}
\usepackage[pdftex,colorlinks=true,linkcolor=blue,citecolor=blue,filecolor=blue,urlcolor=blue]{hyperref}
\usepackage[nameinlink]{cleveref}
\usepackage{subcaption}
\usepackage{ragged2e}
\usepackage{booktabs}
\usepackage[Symbol]{upgreek}
\usepackage{threeparttable}
\usepackage{verbatim}
\usepackage{adjustbox}

\newcommand{\tabincell}[2]{\begin{tabular}{@{}#1@{}}#2\end{tabular}}
\crefformat{figure}{#2Fig.~#1#3}
\crefformat{equation}{#2Eq.~#1#3}
\journal{Astroparticle Physics}

\begin{document}

\begin{frontmatter}



\title{Direct measurement of neutrons induced in lead by cosmic muons
       at a shallow underground site}


\author{}
\address{}

 \author[a,b]{Q.~Du\corref{cor1}}
        \cortext[cor1]{Corresponding author: qiang.du@stu.scu.edu.cn}
 \author[b]{I.~Abt}
 \author[b,c]{A.~Empl}
 \author[b]{C.~Gooch}
 \author[b]{R.~Kneissl}
 \author[a]{S.T.~Lin} 
 \author[b]{B.~Majorovits}
 \author[b,d]{M.~Palermo}
 \author[b]{O.~Schulz}
 \author[e]{L.~Wang}
 \author[e]{Q.~Yue}
 \author[b]{A.J.~Zsigmond}
 \address[a]{College of Physical Science and Technology, Sichuan University, Chengdu 610064, China}
 \address[b]{Max Planck Institute for Physics, Munich 80805, Germany}
 \address[c]{Department of Physics, University of Houston, Houston 77204, USA}
 \address[d]{now at Physics and Astronomy Department, University of Hawaii, Honolulu 96822, USA}
 \address[e]{Department of Engineering Physics, Tsinghua University, Beijing 100084, China}

\begin{abstract}

Neutron production in lead by cosmic muons has been studied
with a Gadolinium doped liquid scintillator detector.
The detector was installed
next to the Muon-Induced Neutron Indirect Detection EXperiment (MINIDEX),
permanently located in the T\"ubingen shallow underground laboratory where
the mean muon energy is approximately 7\,GeV.
The MINIDEX plastic scintillators were used to tag muons;
the neutrons were detected through neutron capture  
and neutron-induced nuclear recoil signals in the
liquid scintillator detector.
Results on the rates of observed neutron captures and nuclear recoils 
are presented and compared to predictions from GEANT4-9.6 and GEANT4-10.3.
The predicted rates are significantly too low for both versions of GEANT4.
For neutron capture events, the observation exceeds the predictions by
factors of
$ 1.65\,\pm\,0.02\,\textrm{(stat.)}\,\pm\,0.07\,\textrm{(syst.)} $ 
and 
$ 2.58\,\pm\,0.03\,\textrm{(stat.)}\,\pm\,0.11\,\textrm{(syst.)} $ 
for GEANT4-9.6 and GEANT4-10.3, respectively.
For neutron nuclear recoil events, 
which require neutron energies above approximately
5\,MeV, the factors are even larger,
$ 2.22\,\pm\,0.05\,\textrm{(stat.)}\,\pm\,0.25\,\textrm{(syst.)} $ 
and 
$ 3.76\,\pm\,0.09\,\textrm{(stat.)}\,\pm\,0.41\,\textrm{(syst.)} $,
respectively.
Also presented is the first statistically significant 
measurement of the spectrum of neutrons induced by cosmic muons
in lead between 5 and 40\,MeV.
It was obtained by unfolding the nuclear recoil spectrum.
The observed neutron spectrum is harder than predicted by GEANT4.
An investigation of the distribution of the time difference between 
muon tags and nuclear recoil signals confirms the validity of the
unfolding procedure and shows that GEANT4 cannot properly describe
the time distribution of nuclear recoil events.
In general, the description of the data is worse for GEANT4-10.3
than for GEANT4-9.6.

\end{abstract}

\begin{keyword}
Neutrons induced by comic muons \sep Neutron spectrum \sep Lead \sep Liquid scintillator

%\PACS 95.30.Cq \sep 14.20.Dh \sep 98.70.Vc \sep 29.40.Mc


\end{keyword}

\end{frontmatter}

\section{Introduction}

Muon-induced neutrons are an important background 
for underground experiments searching for
rare events such as 
neutrinoless double beta decays~\cite{GERDA,PDG}, 
direct dark matter interactions~\cite{dark matter,PDG} or neutrino
interactions in oscillations experiments~\cite{neutrino,PDG}. 
Experiments can sufficiently shield against
neutrons from the radio-impurities in the rock surrounding 
the laboratories, because
the kinetic energy of these neutrons is usually below 10\,MeV and, thus, 
they are efficiently thermalized by low Z materials such as water 
or polyethylene. 
It is more difficult to shield 
against neutrons induced by cosmic muons because
the high-energy muons that penetrate deep into the ground
can produce neutrons with much higher energies.

A particular case are experiments using high-Z materials
like lead, steel and copper for shields~\cite{MAJORANA}
close to the active detector.
The cross sections for muons to generate neutrons in these materials are large
and the neutrons can reach kinetic energies up to several GeV.
These high energy neutrons have a large penetration power and can create
secondary showers with many neutrons reaching the vicinity of
the active detectors.
In addition, the active detector itself can be made out of a high-Z material,
like in the case of Germanium based
experiments to search for neutrinoless double beta decay~\cite{GERDA,MAJORANA}
or dark matter~\cite{CDEX}.
The experiments usually have a muon veto, which is used to reject signals
following the passage of a muon.
However, the neutrons from the showers
can create meta-stable states in the inner structures
or active parts of
an experiment which can decay minutes or hours later and cannot be vetoed
easily. An example is the creation of $^{77m}$Ge from $^{76}$Ge, for which
the decay scheme includes a beta decay with a half-life
of 12 hours~\cite{1802.05040}.

There are four main processes how muons generate neutrons inside matter: 
\begin{itemize}
\item  muon-nuclear deep inelastic scattering; 
\item  photo-nuclear reactions, i.e.\ Bremsstrahlung photons 
       induce photo-disintegration; 
\item  hadronic inelastic scattering, i.e.\ muon-induced secondary 
       hadrons cause hadron-induced spallation; 
\item  $\upmu^{-}$ capture, i.e.\ nuclei excited by $\upmu^{-}$ capture 
       release neutrons. 
\end{itemize}
Muon capture is only relevant in shallow underground sites 
with a depth of $\lesssim$\,100\,m water equivalent~($mwe$)~\cite{Boulby-1}. 

The production of
muon-induced neutrons has been investigated for different materials 
and at different depths for many years~\cite{H. M. Kluck}. 
In this paper, the focus is on lead as a target material.
Details about some selected experiments on lead 
are listed in Table~\ref{tab:review}.

\begin{table}[t]
\centering
\caption{\label{tab:review} Underground experiments on 
                            neutrons induced by cosmic muons in lead}
        \begin{tabular}{@{\extracolsep{\fill}}rcl}
		\toprule 		
		\tabincell{c}{Depth \\ $[$mwe$]$}
                & \tabincell{c}{Lead thick- \\ness [cm]}    &  Reference     \\
			\toprule
			  12  			             & 15       &    G. V. Gorshkov(1974)\cite{Gorshkov1974}     \\
%			\hline
			  16                         & 50       &    I. Abt (2017)\cite{Iris2017} and this work \\
%			\hline			
			  20			             & 7.6      &    M. F. Crouch (1952)\cite{Crouch1952}\cite{Annis1954}             \\
%			\hline
			  40  			             & 10       &    G. V. Gorshkov (1971)\cite{Gorshkov1971a}                    \\
%			\hline			
			  58			             & ---$^{a}$      &    Holborn (1965)\cite{Holborn}                          \\
%			\hline
			  60    			         & 10       &    L. Bergamasco (1970)\cite{Bergamasco1970}                   \\
%			\hline			
			  80			             & 10       &    G. V. Gorshkov (1971)\cite{Gorshkov1971a}                    \\
%			\hline
			 110    			         & 10       &    L. Bergamasco (1970)\cite{Bergamasco1970}                   \\
%			\hline			
			 150			             & 10, 16   &    Gorshkov (1968\cite{Gorshkov1968}, 1971\cite{Gorshkov1971}) \\
%			\hline
			 800  			             & 10       &    G. V. Gorshkov (1971)\cite{Gorshkov1971a}                    \\
%			\hline
			 2850    			         & ---$^{b}$      &    Boulby (2008\cite{Boulby-1}, 2013\cite{Boulby-2})       \\
%			\hline
			 4300  			             & 35       &    L. Bergamasco (1973)\cite{Bergamasco1973}                   \\
%			\hline
			 4800  			             & 10       &    LSM (2013)\cite{H. M. Kluck}                      \\				 				 		 			 
		\bottomrule
        \end{tabular}
\flushleft{\footnotesize 
$^{a}$ The lead was mixed with rock and the neutron spectrum 
was measured with low statistical significance. \\
$^{b}$ The target was the whole lead shielding 
system of the ZEPLIN-II and ZEPLIN-III experiments.\\
}
\end{table}

Of the experiments previous to this work listed in Table~\ref{tab:review},
only the experiment at the Holborn underground laboratory~\cite{Holborn} 
has published a measurement of the spectrum of the neutrons 
induced by cosmic muons, albeit with low statistical significance. 
In addition, the target was composed of a mixture of lead and rock and
no corresponding  simulation was available. 
Complementary to the underground experiments on neutrons induced by 
cosmic muons,
two accelerator based experiments measured  muon-induced 
neutron spectra in lead.
These were the E665 experiment at Fermilab using a 470\,GeV 
muon beam~\cite{E665} and the NA55 experiment at CERN using
a 190\,GeV muon beam~\cite{NA55}. The results of the NA55 experiment 
were compared to GEANT4-8.0 simulations and significant discrepancies 
in neutron multiplicities and angular distributions 
were found~\cite{M. G. Marino}. An updated experiment using a 160\,GeV 
muon beam has recently been performed~\cite{Y. Nakajima}.

In this paper, a detailed investigation of neutrons induced by cosmic muons
in lead is presented. 
The experiment was performed in the shallow underground laboratory at the
University of T\"ubingen. Thus, the results cannot be directly used to
predict the background in deep underground laboratories. They can, however,
be used to evaluate the Monte Carlo (MC) programs used to make such predictions.

To measure the spectrum of muon-induced neutrons,
a 28-liter Gadolinium doped (0.5\,\% in weight) liquid scintillator (Gd--LS) 
detector was installed next to the MINIDEX
(Muon-Induced Neutron Indirect Detection EXperiment)~\cite{Iris2017} setup.
The details of the experimental setup and the detector concept 
are introduced in Section~\ref{sec:setup}. 
The energy calibration is described in Section~\ref{sec:energy calibration}. 
The selections of events and their processing are presented
in Section~\ref{sec:data analysis}. The details of the MC 
simulations are described in Section~\ref{sec:simulation}. 
The results are given in Section~\ref{sec:results}. 
A discussion on the MC performance and a summary are presented in the 
last two Sections.

\section{Experimental setup and detector concept}
\label{sec:setup}
\begin{figure}[t]
\centering
\includegraphics[width=0.48\textwidth]{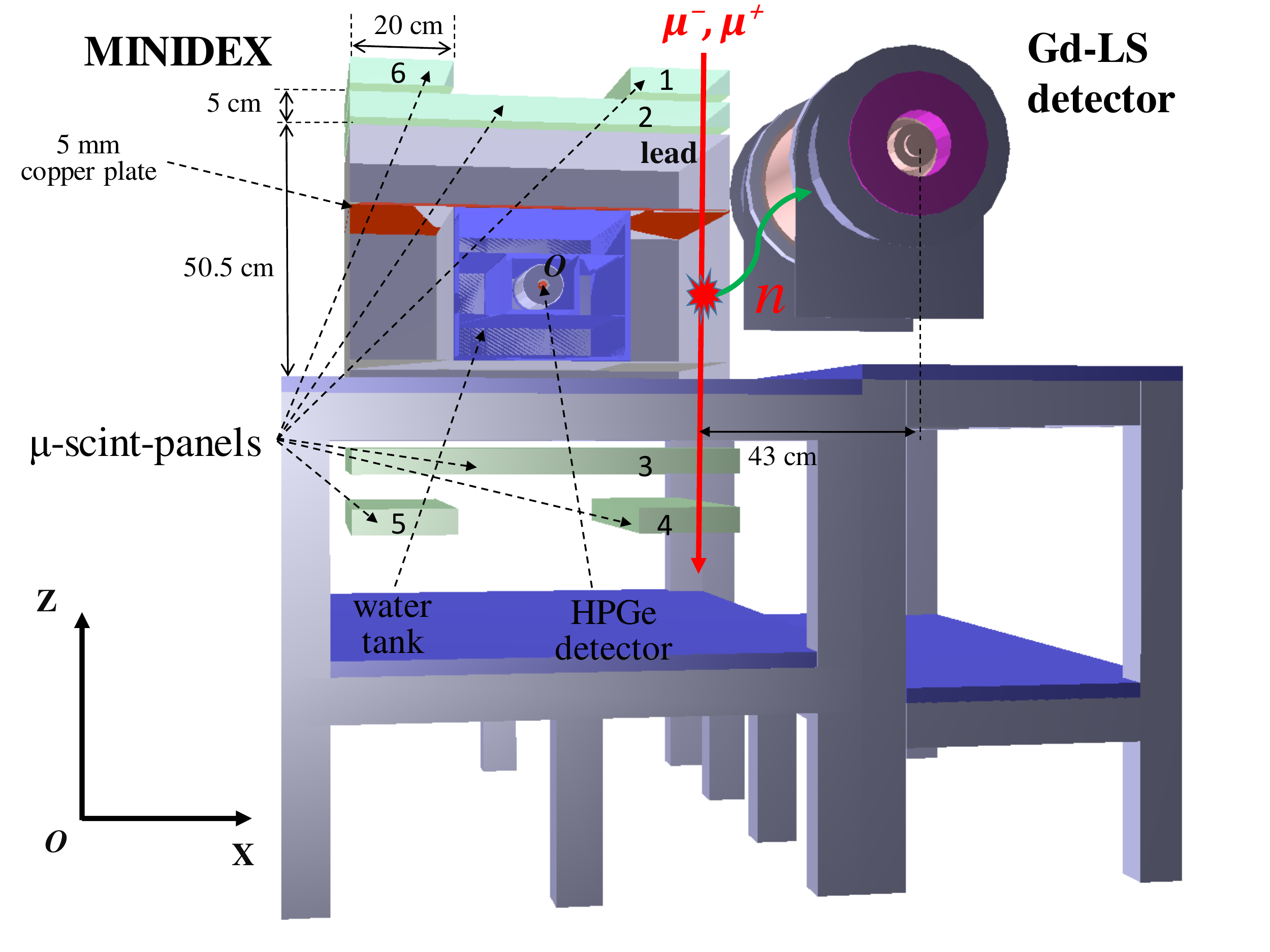}
\caption{\label{fig:setup} The MINIDEX setup together with
the Gadolinium doped liquid scintillator (Gd--LS) detector. 
The six muon scintillator panels 
($\upmu$-scint-panels) were used to independently identify cosmic muons passing 
through the left-side and right-side lead walls. 
Signals in the four muon scintillator panels labeled 1, 2, 3 and 4 
were required to select muon-induced signals in the Gd--LS detector. 
A steel box enclosing the Gd-LS detector is not shown.}
\end{figure}

The Gd--LS detector was placed next to one of the lead walls of
the MINIDEX in the underground laboratory
in T\"ubingen. 
The nominal vertical depth of the site
was given~\cite{Depth} as 16\,$mwe$; the cover consists 
predominantly of soil.
According to simulation,
which is described in Section~\ref{sec:simulation},
the average energy of the cosmic muons entering the laboratory
was $\approx 7$\,GeV~\footnote{The technical drawings and
  the result of the simulation indicate that the value
  of 16\,$mwe$ is an overestimate and that the overburden more closely
  corresponds to 12\,$mwe$, see also Section~\ref{sec:simulation}.}.
The geometry of the MINIDEX setup 
plus the Gd--LS detector is shown in Fig.~\ref{fig:setup}. 
A detailed description of MINIDEX is given in \cite{Iris2017}. 
Here, only components relevant for the analysis 
of the data from the  Gd--LS detector data are described.

In the standard MINIDEX analysis, neutrons were identified with
two high-purity germanium (HPGe) detectors embedded in a water tank. 
These were used to detect 2.2\,MeV $\upgamma$-rays 
emmited after neutron capture in water. 
The dimensions of the water tank were $35\times 55 \times 30$\,cm$^{3}$ 
with a $15\times 55 \times 10$\,cm$^{3}$ central cavity in which 
the HPGe detectors were inserted. 
The water tank and the HPGe detectors were fully surrounded by a lead castle, 
used both as a shield for the HPGe detectors and as the target for muons 
to generate neutrons. 
The thickness of the lead castle was 15\,cm at the top supported by a 
5\,mm thick copper plate, 
5\,cm at the bottom and 5\,cm for the two side walls  
at right angle to the HPGe and Gd--LS detectors which 
are not shown in Fig.~\ref{fig:setup}. 
The two lead walls parallel to the Gd--LS detectors were 
20\,cm thick. These two walls are hereafter referred to as the left-side and 
right-side lead walls where the right-side wall is next to the Gd--LS detector. 

\begin{figure}[t]
\centering
\includegraphics[width=0.48\textwidth]{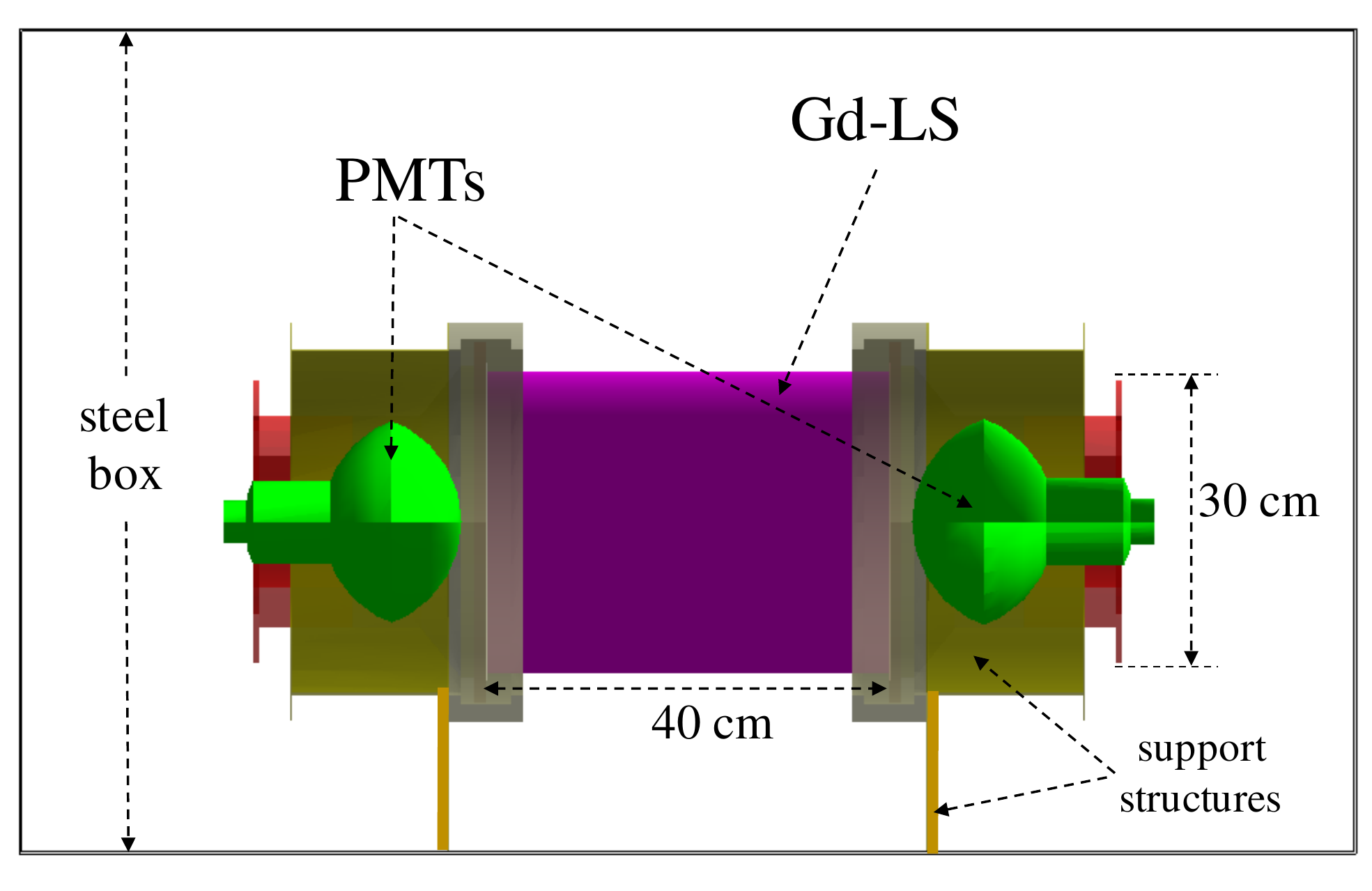}
\caption{\label{fig:detector structure} Schematic of the Gd--LS detector. 
The Gd-LS type EJ-335 was contained in a 8\,mm thick 
tempered borosilicate glass cylinder with a diameter of 30\,cm diameter 
and a length of 40\,cm. Two 8--inch photomultiplier tubes (PMTs) were 
connected to the two flat end-surfaces of the glass cylinder 
with light guides. The active detector was located inside a 
light-tight steel box.}
\end{figure}

A schematic of the Gd--LS detector is shown in 
Fig.~\ref{fig:detector structure}. 
Organic liquid scintillator (EJ-335) was held
inside a 8\,mm thick tempered borosilicate glass cylinder with 
a diameter of 30\,cm  and  a length of 40\,cm. 
The EJ-335 was produced by the Eljen Technology company 
and was loaded with 0.5\% Gadolinium by weight~\cite{EJ-335}. 
Two 8--inch photomultiplier tubes (PMTs) were connected to 
the two flat end-plates of the glass cylinder with light guides. 
The active detector was protected inside a light-tight steel box. 
The distance between the center of the Gd--LS detector and the center of
the right-side lead wall was 43\,cm. 

Data was taken together with  MINIDEX from January to July 2016.
The data acquisition was run without trigger, i.e. the data stream
from all components was recorded continuously.
The association of neutron signals to muons was done offline
with the plastic scintillator panels of MINIDEX being used as
a muon tagger.
The dimensions of the muon scintillator panels  
were $75\times 65\times 5$\,cm$^{3}$ for the 
two big panels (2 and 3 in Fig.~\ref{fig:setup}) 
and $20\times 65\times 5$\,cm$^{3}$ for the 
four small panels (1, 4, 5 and 6 in Fig.~\ref{fig:setup}). 
The two big panels fully covered 
the top and the bottom of the lead castle. 
For the analysis presented here, only muons 
passing through the full lengths (50\,cm) of the lead wall 
next to the Gd--LS detector were considered.
They were identified by coincidences of signals in the four 
panels 1, 2, 3 and 4; this will be described in detail in
Section~\ref{sec:muon condition}.

\begin{figure}
\centering
\includegraphics[width=0.48\textwidth]{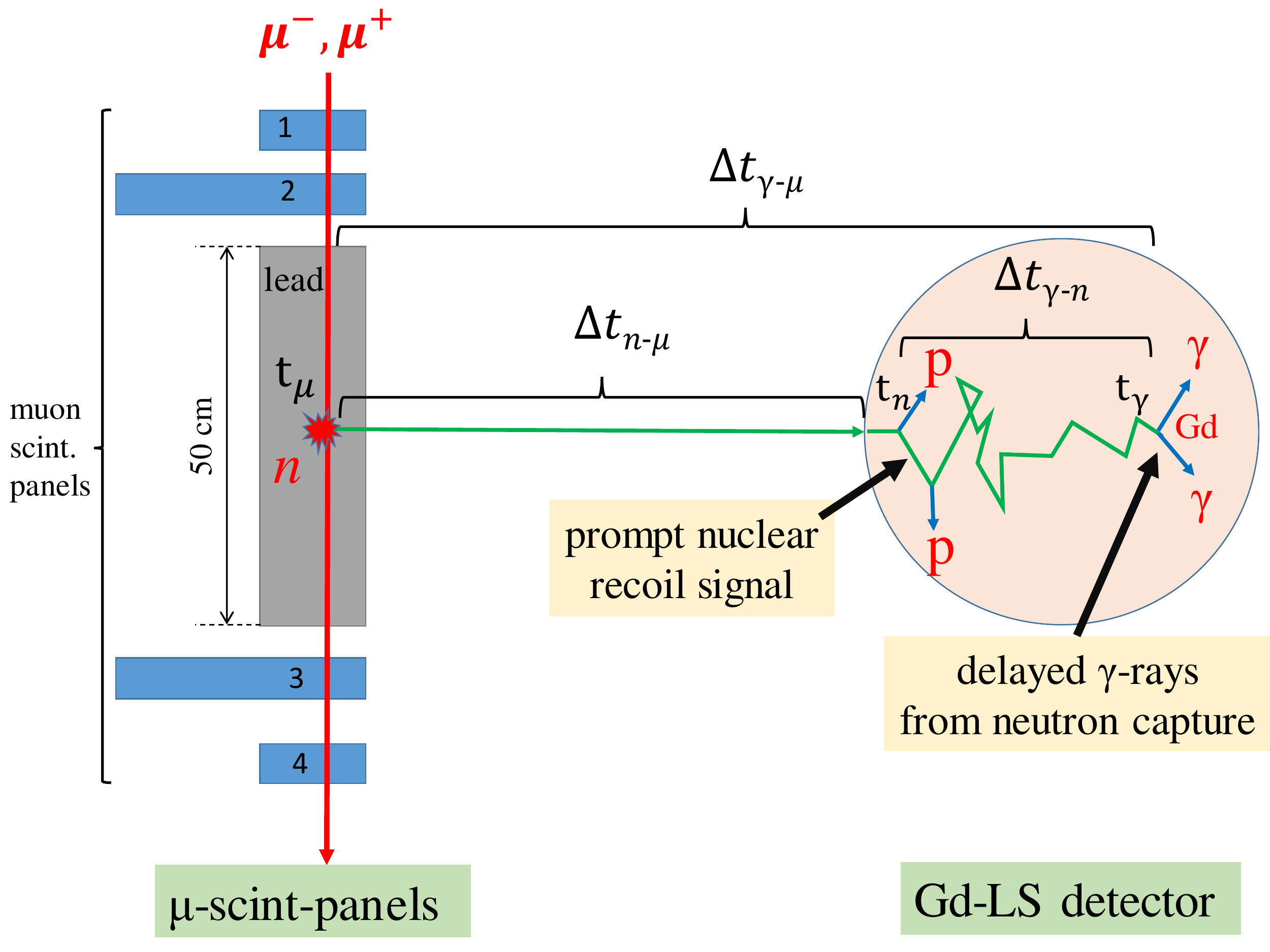}
\caption{\label{fig:detection principle} The principle of detecting 
muon-induced neutrons in the Gd--LS detector.}
\end{figure}

Figure~\ref{fig:detection principle} illustrates the detection 
of neutrons, which were created by cosmic muons in the lead, 
in the Gd--LS detector. 
If a muon passage through the right-side lead wall was tagged, 
following neutron signals were searched for.
Neutrons created in the lead by muons directly or 
in a muon-induced shower at time $t_{\upmu}$ had a finite probability
to propagate into the Gd--LS detector. 
The range of typical times of flight $\Delta t_{n\textrm{-}\upmu}$ 
is tens to hundreds of nanoseconds. 
Neutrons with a sufficiently high energy will first 
create a signal from a prompt nuclear recoil at time $t_{n}$ 
through (multiple) elastic scattering on the nuclei of 
the liquid scintillator (proton or carbon).
Then, after thermalization, neutrons will diffuse 
for a few microseconds ($\sim$\,7\,$\upmu$s) until they are captured 
by Gd. A neutron capture is followed by a rapid ($<$ns) release of 
several $\upgamma$-rays with a total energy of around 8\,MeV 
at time $t_{\upgamma}$. 
This results in a $\upgamma$-signal delayed with respect to 
the nuclear recoil signal by $\Delta t_{\upgamma\textrm{-}n}$
and to the muon signal by $\Delta t_{\upgamma\textrm{-}\upmu}$. 

The unique combination of the muon signal, 
the prompt nuclear recoil signal and the delayed $\upgamma$-signal, 
as well as the characteristic 
times $\Delta t_{n\textrm{-}\upmu}$, 
$\Delta t_{\upgamma\textrm{-}n}$ and $\Delta t_{\upgamma\textrm{-}\upmu}$ 
provide a highly background suppressed measurement of neutrons 
induced in lead by muons.
In most muon-induced neutron experiments, the detectors are located below 
or above the targets. In this case, the Gd--LS detector was placed at 
the side of the target which is of great advantage as the
detection of the neutrons was not disturbed by the muons producing them.

\section{Energy calibration and threshold}
\label{sec:energy calibration}
The energy calibration of the Gd--LS detector was performed
using the detector response to a variety of $\upgamma$ rays
and to minimum ionizing cosmic muons (m.i.p.s.).  
The gamma rays originated from
$\upgamma$ lines of a $^{60}$Co and a $^{228}$Th source, 
the 2.2\,MeV $\upgamma$-line from de-excitation after neutron capture 
in hydrogen and the 4.438\,MeV $\upgamma$-line from an AmBe neutron source.

The response of the detector to $\upgamma$ rays from the calibration sources
and muons was simulated by implementing the detector setup in GEANT4~\cite{geant4}. 
The energy resolution of the detector was limited and had to be determined 
together with the energy scale.
It was modeled as a Gaussian 
$\sigma=\sqrt{a\cdot E}$~\cite{Boulby} with a variable
parameter $a$ to be adjusted.
The resolution parameter $a$ was determined to be 30\,keV by matching 
the observed and simulated widths of the peaks.
The visible line energies were also fitted.
Due to the limited resolution they are shifted downwards from
the nominal $\upgamma$ line energies because part of the
Compton shoulders could not
be separated from the observable peaks.
All calibration sources and the nominal and visible energies of 
the $\upgamma$ lines are listed in Table~\ref{tab:energy calibration}. 

\begin{figure*}
\centering
\begin{subfigure}{0.48\textwidth}
\includegraphics[width=\linewidth]{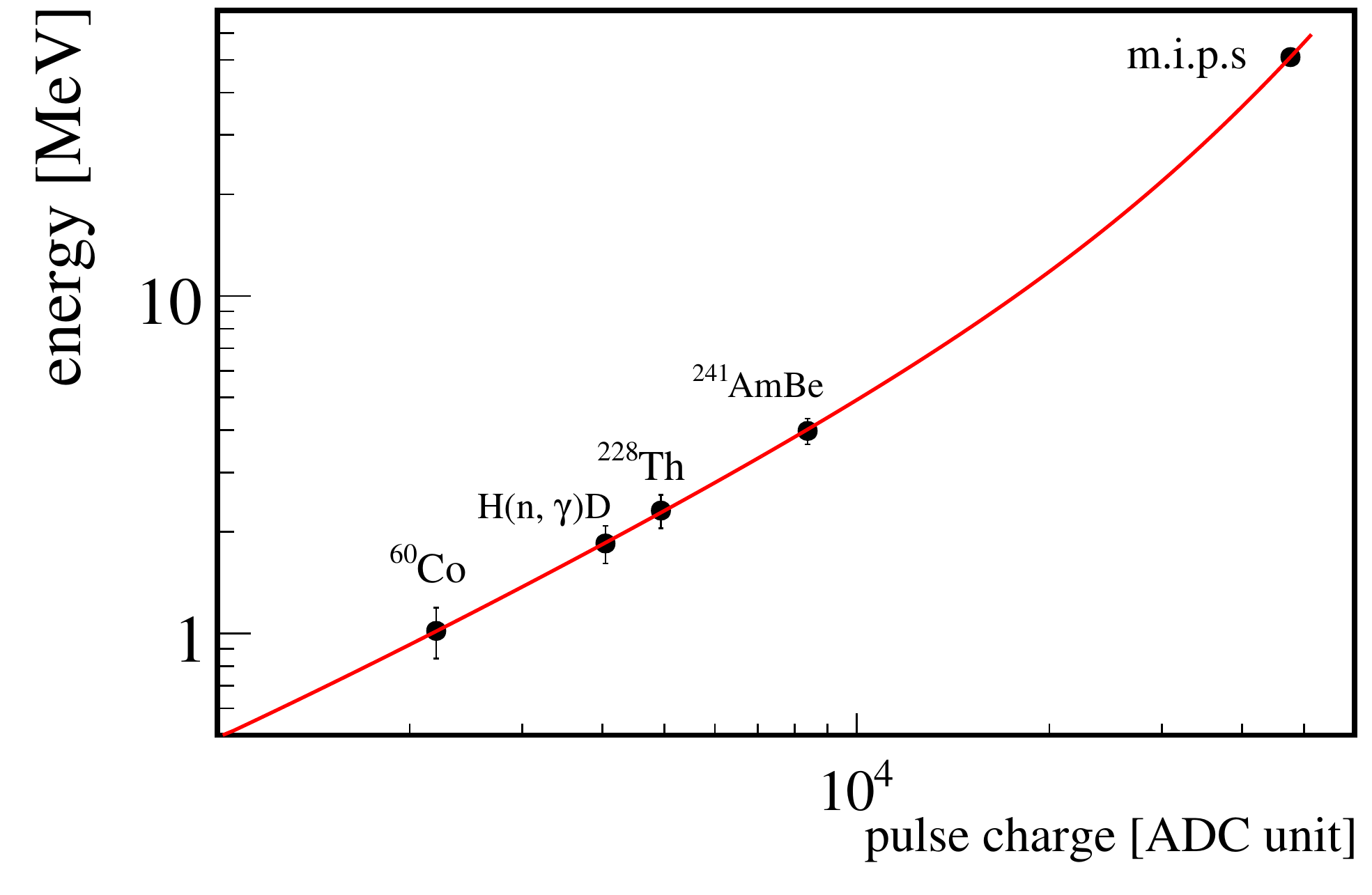}
\caption{\label{fig:calibration function}}
\end{subfigure}
\hfill
\begin{subfigure}{0.48\textwidth}
\includegraphics[width=\linewidth]{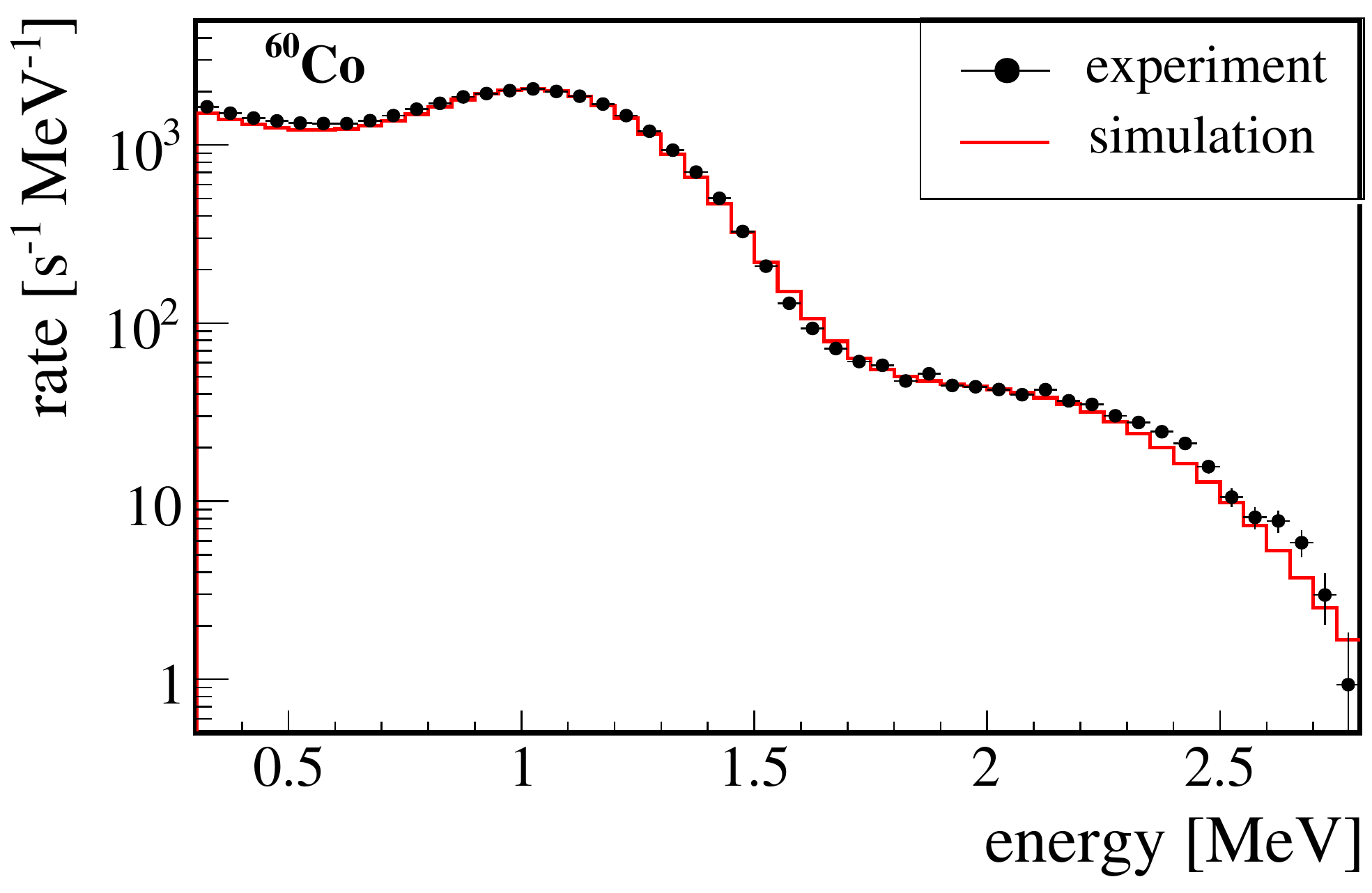}
\caption{\label{fig:Co60 spectrum}}
\end{subfigure}
\begin{subfigure}{0.48\textwidth}
\includegraphics[width=\linewidth]{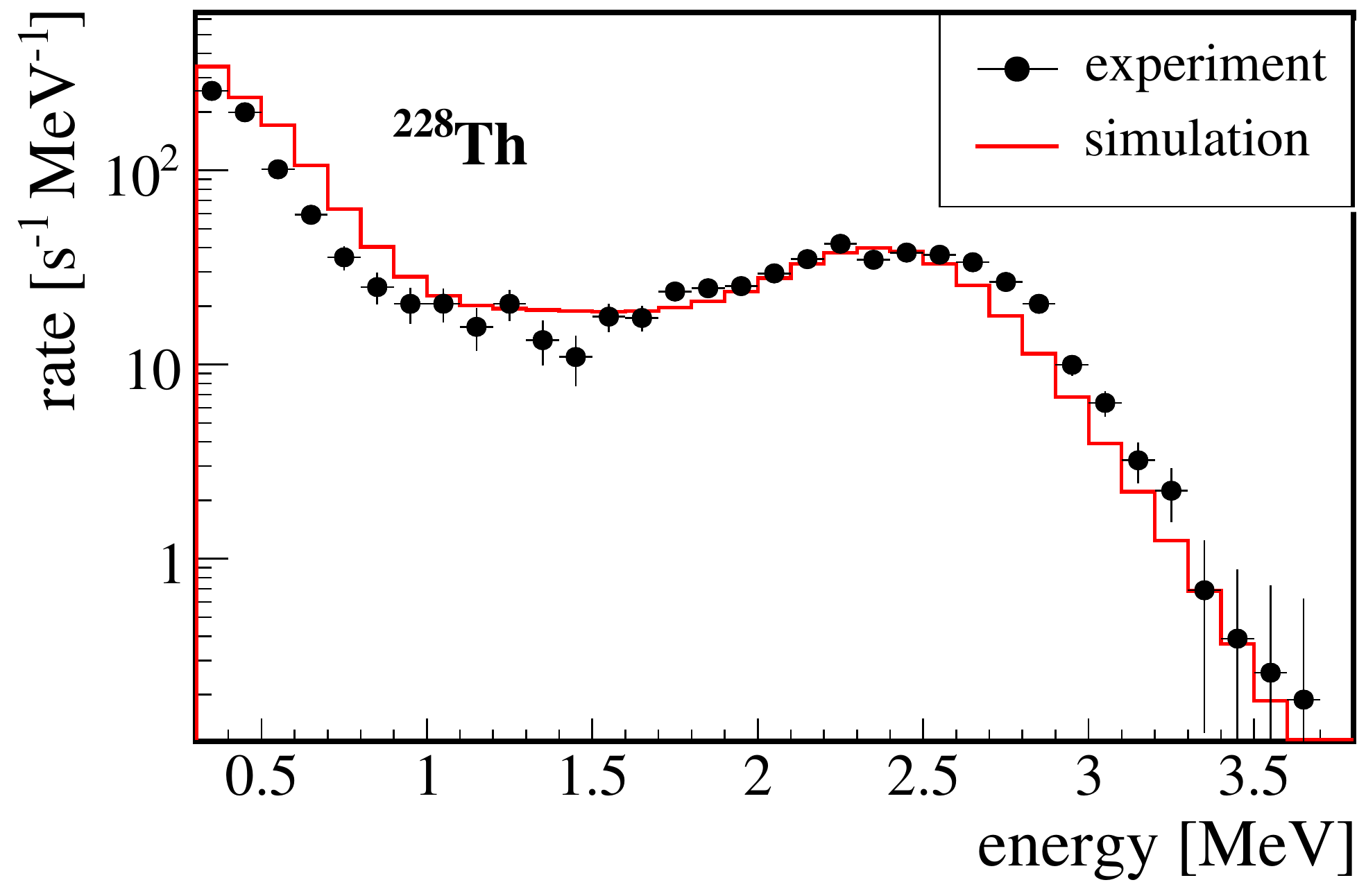}
\caption{\label{fig:Th228 spectrum}}
\end{subfigure}
\hfill
\begin{subfigure}{0.48\textwidth}
\includegraphics[width=\linewidth]{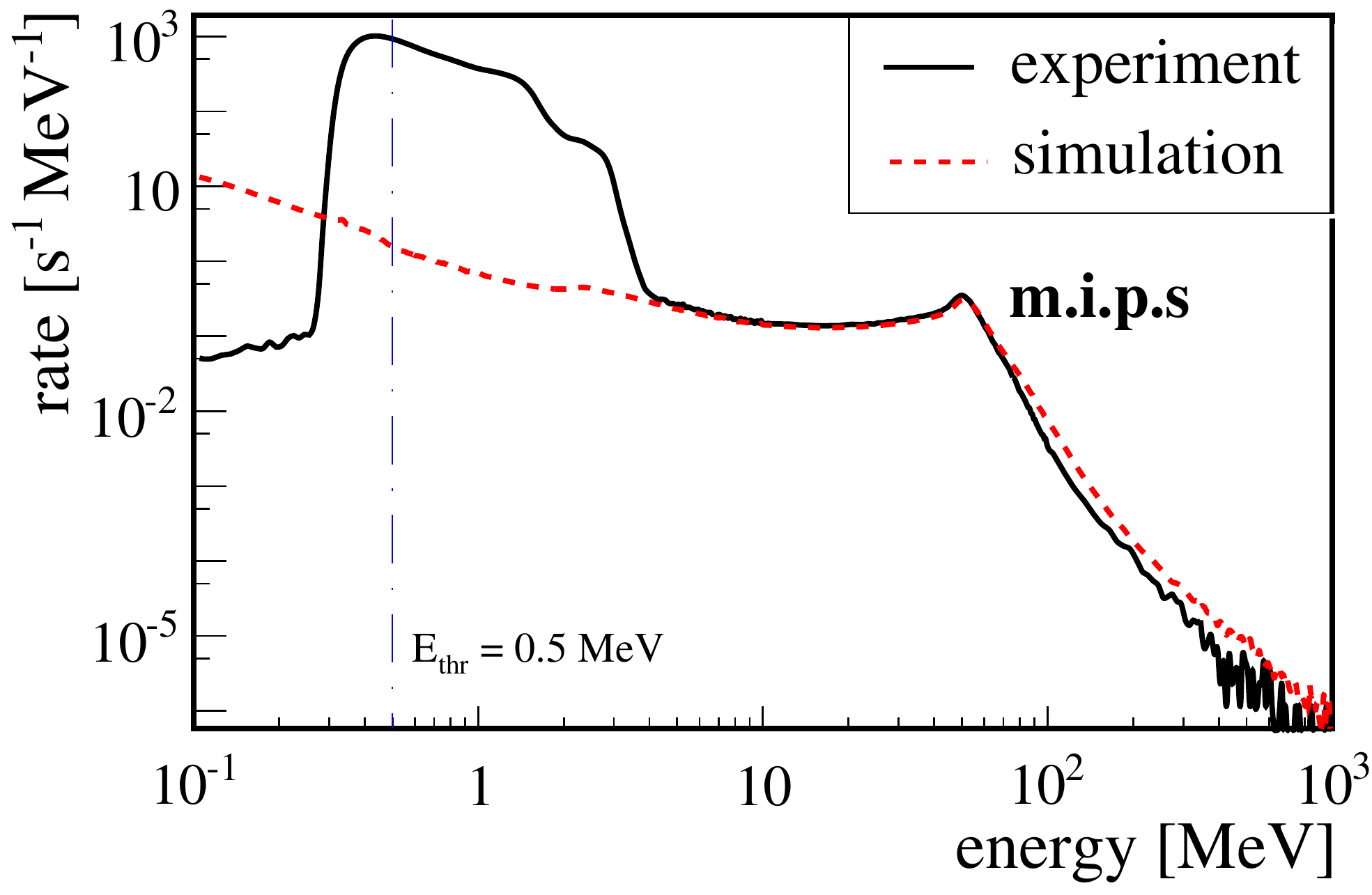}
\caption{\label{fig:total spectrum}}
\end{subfigure}
\caption{\label{fig:energy calibration} Energy calibration. 
  (a) Visible line energy versus so-called pulse charge as recorded
   for the PMTs as response to the sources listed in 
   Table~\ref{tab:energy calibration}. The solid curve represents
   the result of a fit of a third-order polynomial to the points.
   The two lines from $^{60}$Co are not resolved.
   The error bars represent the resolutions calculated with the
   parameter $a = 30$\,keV.
   This resolution parameter was used for all simulated distributions. 
   Calibrated measured and simulated energy spectra for 
  (b)$^{60}$Co,
  (c)$^{228}$Th.
  (d) Calibrated measured energy spectrum observed in the laboratory and 
   predicted spectrum (GEANT4-9.6) of the energy deposited by cosmic muons 
   and their secondary particles in the Gd--LS detector.
   The peak at 51\,MeV is due to the passage of minimum ionizing 
   particles. 
   The simulated and the measured spectra are normalized to their corresponding 
   lifetimes. 
}
\end{figure*}

The so-called pulse charges 
were calculated in ADC units
by combining the integrated pulses of the two PMTs. 
The peaks in the resulting spectra were fitted to obtain the
visible line energies in ADC units. These 
were calibrated against the visible line values obtained from
equivalent fits to the simulated energy spectra, see 
Table~\ref{tab:energy calibration}. 
The resulting calibration points are shown in 
Fig.~\ref{fig:calibration function}, together with a fit
of a third order polynomial function to these points.
The result of this fit was used as the energy calibration function.
 
\begin{table}
\centering
\caption{\label{tab:energy calibration} 
Radiation sources used for the energy calibration. 
For the $\upgamma$ lines, the nominal line energies and the visible
line energies together with their resolutions calculated
with the parameter of $a = 30$\,keV are listed. 
The two lines from $^{60}$Co are not resolved.
For m.i.p.s, only the visible line energy from the GEANT4 simulation 
is available.}
        \begin{tabular}{@{\extracolsep{\fill}}ccc}
		\toprule 		
		    \tabincell{c}{Radiation\\Source}  & \tabincell{c}{Nominal Line \\Energy [MeV]}  &  \tabincell{c}{Visible Line \\ Energy [MeV]} \\
			\toprule
			  $^{60}$Co  			                & 1.17/1.33                         & $1.02\pm0.17$   
\\
%			\hline			
			  H(n, $\upgamma$)D	                    & 2.2                               & $1.85\pm0.24$                \\
%			\hline
			  $^{228}$Th  			                & 2.615                             & $2.31\pm0.26$                \\
%			\hline			
			  $^{241}$AmBe			                & 4.438                             & $3.98\pm0.35$                \\
%			\hline
			  m.i.p.s                               &  ---                              & $51.0\pm1.2$                \\
		\bottomrule
        \end{tabular}
\end{table}

The calibrated data from the sources are shown together with the simulated spectra
for $^{60}$Co and $^{228}$Th in
Figs~\ref{fig:Co60 spectrum} and~\ref{fig:Th228 spectrum}
respectively. 
The agreement is quite good.

The calibrated spectrum in the laboratory 
as measured with the Gd--LS detector is shown 
together with the simulated spectrum of energy deposited by cosmic muons 
and their secondary particles in Fig.~\ref{fig:total spectrum}. 
The details of the simulation of cosmic muons
passing through the overburden to the underground laboratory
are discussed in Section~\ref{sec:simulation}.
Both simulated and measured spectra are normalized to their corresponding 
lifetimes. 
The agreement is quite good above 4\,MeV. 
The discrepancy below 4\,MeV is due to environmental radioactivity
which was not simulated.

For the event selection, a threshold of $E_{thr}$\,=\,0.5\,MeV 
was set offline. Figure~\ref{fig:total spectrum} shows that this
was safely above the hardware threshold of the Gd--LS detector which was
$\approx$0.3\,MeV.

The energy calibration and threshold are based on the unit MeV.
Neutron interactions deposit energies which are not as directly related
to incident energies as is the case for $\upgamma$ or muon interactions.
Therefore, from here on, the unit MeV$_{ee}$, where $ee$ stands for
electron equivalent, is used for neutron interactions, for which 
quenching is taken into account. 

\section{Event Selection}
\label{sec:data analysis}

The neutrons from muon-induced interactions are identified
by signals recorded after the observation of a muon tag.
The sample of events with the least background contamination 
are called nuclear recoil events.
They comprise three signals, a muon tag from the muon scintillator panels,
a prompt signal from a nuclear recoil due to a neutron scattering off
a nucleus in the Gd--LS detector,
and a delayed $\upgamma$ signal from a neutron capture in the
Gd--LS detector. 
The selection of these events demonstrates the capabilities of the
detector best and therefore is described first.

A less restrictive and larger sample is composed of so called
capture events, which are events with only a muon tag and 
a delayed capture signal. This sample contains also neutrons with
energies too low to create a detectable nuclear recoil signal.

\subsection{Muon tagging}
\label{sec:muon condition}

\begin{figure*}
\centering
\begin{subfigure}{0.48\textwidth}
\includegraphics[width=\linewidth]{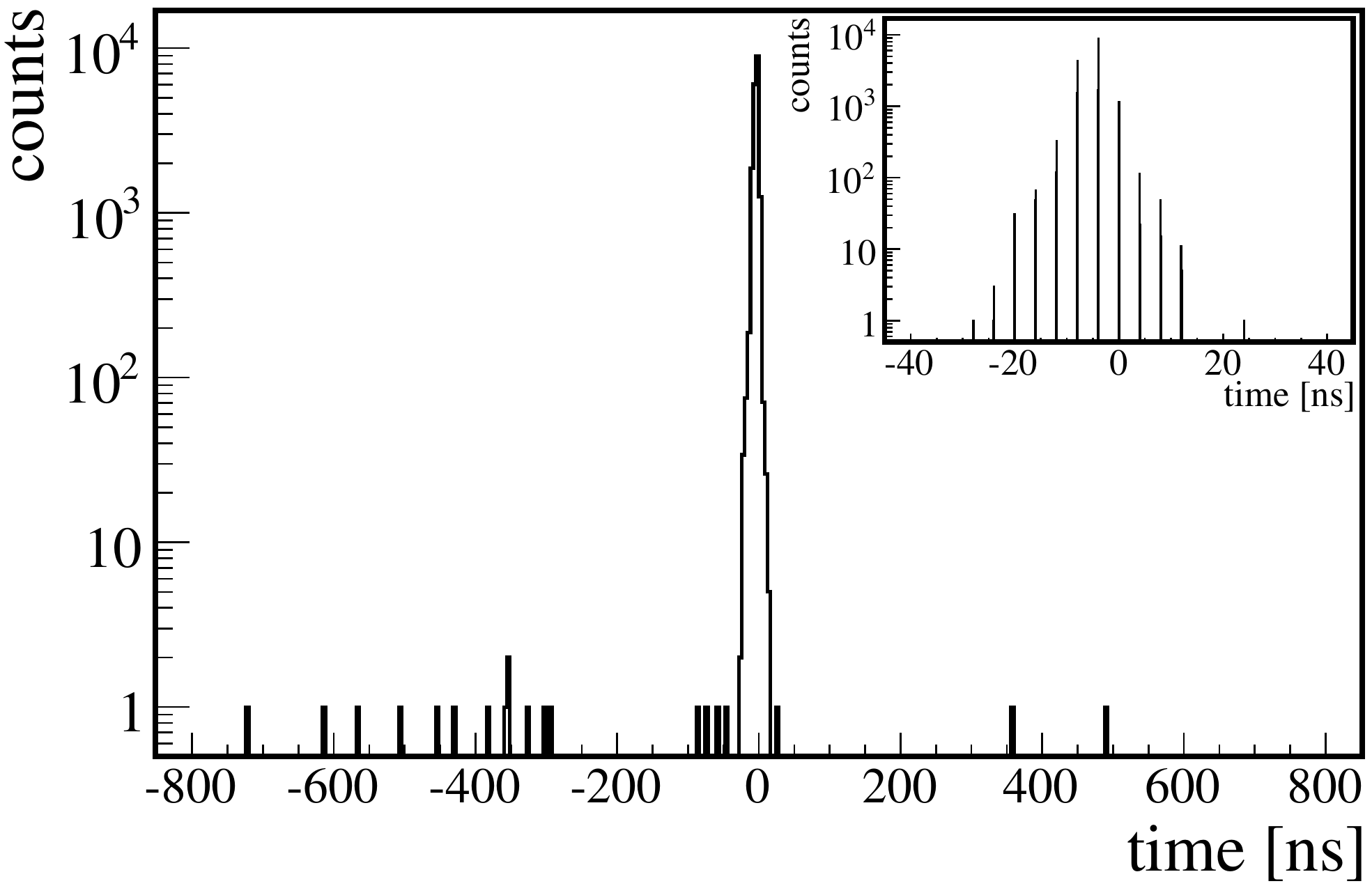}
\caption{\label{fig:muon time}}
\end{subfigure}
\hfill
\begin{subfigure}{0.48\textwidth}
\includegraphics[width=\linewidth]{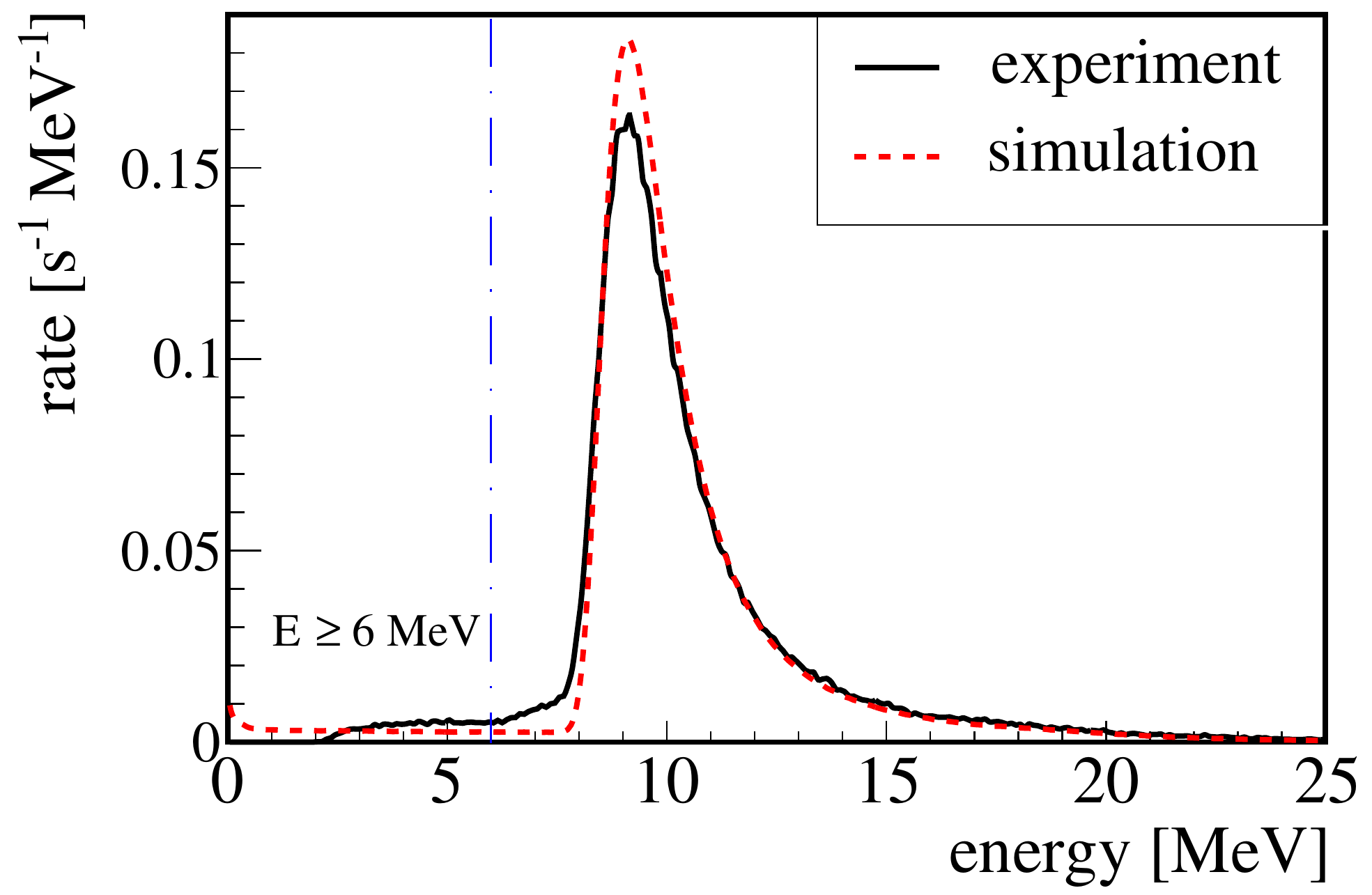}
\caption{\label{fig:muon spectrum}}
\end{subfigure}
\caption{\label{fig:muon selection} Selection of muon tags.
  (a) Distribution of measured time differences for muon scintillator panel number~1 and number~4 for muon tags;
  the ADC sampling time is 4\,ns, see insert.
  (b) Spectrum of the energy deposited in 
  muon scintillator panel number~1 for muon tags. Also shown is the spectrum as predicted
  by simulation (GEANT4-9.6). The integral of the simulated spectrum 
  was normalized to the integral of the measured spectrum.}
\end{figure*}

The passage of a muon through the right-side lead wall was tagged by
coincident signals in the four muon scintillator panels (1, 2, 3, 4), 
see Fig.~\ref{fig:detection principle}.
The time difference between
each pair of panels had to be below 30\,ns.
This was motivated by the measured time differences for 4--fold
coincidences which are shown in Fig.~\ref{fig:muon time}.
The efficiency of this timing cut is 100\,\% while the
background for 4--fold coincidences can be neglected.
The contribution of random coincidences was estimated
from the single panel rates to be  $7\times10^{-6}$.
The passage of multiple muons and coincidences due to secondary 
particles were not discriminated against.

Figure~\ref{fig:muon spectrum} shows the distribution
of the energy deposited in muon scintillator panel number~1
for such 4--fold coincidences;
also shown is a corresponding simulation, see Section~\ref{sec:simulation}
for details.
The integral of the simulated spectrum was normalized to the integral of the
measured spectrum in order to compare the shapes. 
The peak in the spectrum, associated with the passage 
of minimum ionizing particles, 
was at $\approx 10$\,MeV. 

The observed distributions for panels number~1 and~4 had a 
tail towards lower energies with
a kink at 6\,MeV. The simulated distributions also have a tail
towards lower energies associated with muons which only grazed 
either panel~1 or panel~4. However, the observed distributions 
indicate that
the scintillator panels additionally had some inefficiencies for
grazing muons. 
This caused entries in the shoulder above 6\,MeV.
Below 6\,MeV, the observed distribution flattens out but this is
already close to the threshold. 
Thus, for a muon tag, it was required that the energy deposited in 
all four scintillator panels was above 6\,MeV.
According to simulation, the requirement of this minimum energy for all
four scintillator panels led to an efficiency of 94.6\%.

The lifetime of the experiment was 151.6 days.
In the data, 7.30 million muon tags were identified. 
The number of muon tags would have increased by 1.1\%, if 
the energy cut would have been lowered to 5\,MeV. 
It would have increased by 0.01\%, if only a coincidence between 
muon scintillator panels~1 and 4 would have been required; this shows that
the extra requirement to have signals in the large panels does not
significantly diminish the efficiency for muons
while reducing random coincidences to an insignificant level.  

\subsection{Prompt signals in the Gd--LS detector as nuclear recoil candidates}
\label{sec:nuclear recoil}

The first signal in the Gd--LS detector following a muon tag in a predefined
window was selected as a candidate for a signal from a nuclear recoil.
As such a signal is expected to arrive on the scale of tens
to hundreds of nanoseconds,
it is called a prompt signal.
The position of this time window was established according to simulation,
see Section~\ref{sec:simulation} for details, and according to
delays due to the experimental setup.

\begin{figure}[t]
\centering
\includegraphics[width=0.48\textwidth]{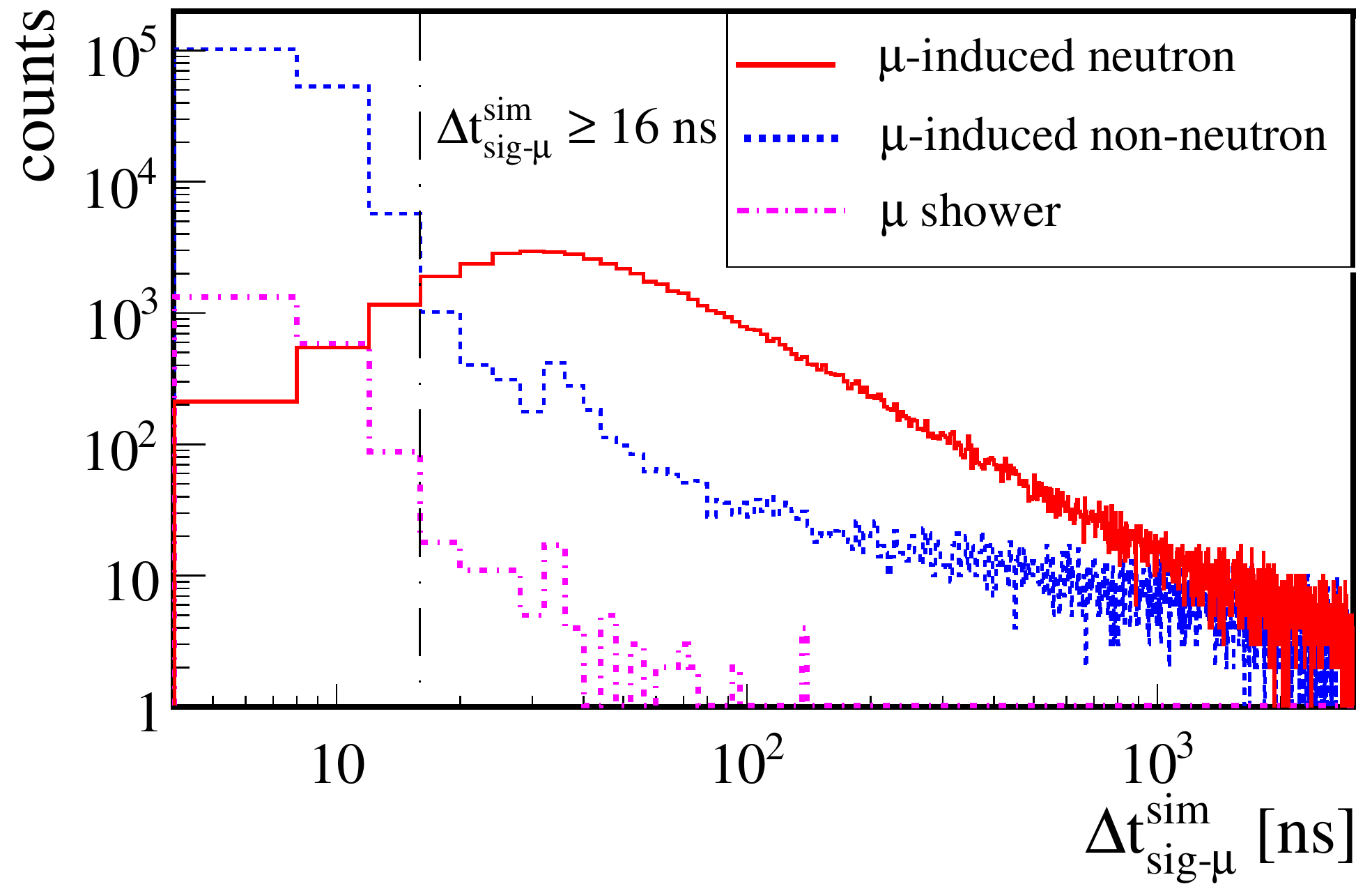}
\caption{\label{fig:simulated time interval to muon trigger} 
Simulated (GEANT4-9.6) time
between the first signal in the Gd--LS detector and
a muon tag, $\Delta t_{\textrm{sig-} \upmu}^{\textrm{sim}}$, for 
$\upmu$-induced neutrons, $\upmu$-induced other particles (non-neutrons)
and secondary particles from showers induced by the muon
before it entered the underground laboratory (shower).}
\end{figure}

The simulated distributions of the time, 
$\Delta t_{\textrm{sig-} \upmu}^{\textrm{sim}}$, between a muon tag and a 
signal in the Gd--LS detector are shown in 
Fig.~\ref{fig:simulated time interval to muon trigger} for
the individual contributions from 
muon-induced neutrons,
muon-induced particles other than neutrons 
and secondary particles in muon-induced showers.
The different origins of the particles led
to different momentum distributions as different types of
particles undergo different types of interactions on their 
paths from the right-side lead wall to the Gd--LS detector.
As a result, the muon-induced neutrons take longer to reach 
the Gd--LS detector than other particles. 
As indicated in Fig.~\ref{fig:simulated time interval to muon trigger},
the requirement of a time delay of 16\,ns rejects
most of the muon-induced non-neutron background, 
while maintaining a neutron detection efficiency of 97\,\%. 

\begin{figure}
\centering
\includegraphics[width=0.48\textwidth]{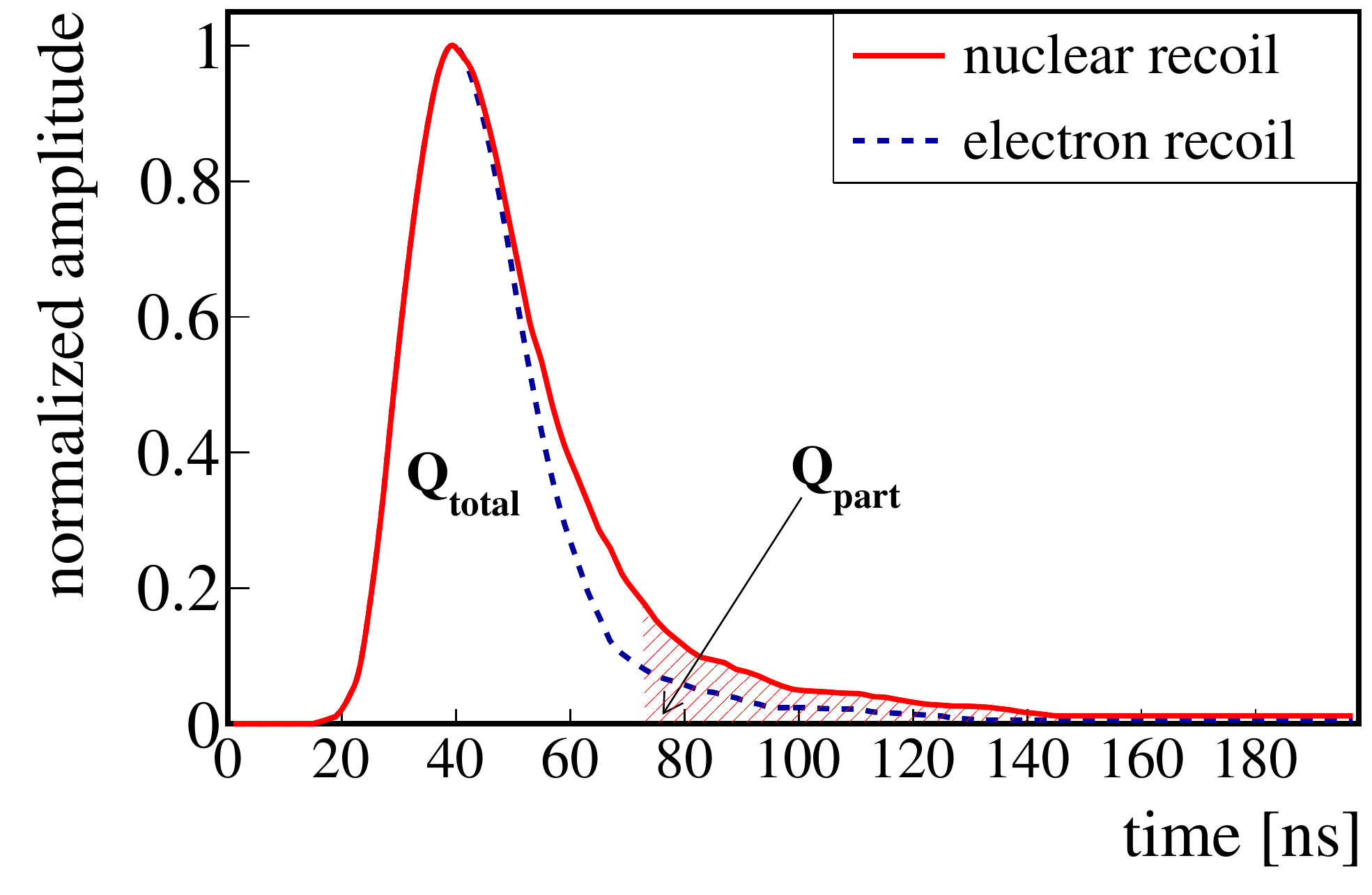}
\caption{\label{fig:pulse shape} Average waveform of neutron-induced nuclear 
recoil signals. Also shown are electron recoils induced 
by $\upgamma$, e$^{-}$ or muon radiation. The data were obtained with an
identical Gd--LS detector; plot taken from~\cite{Qiang2017}. 
The parameter Q$_{total}$ denotes the total integral of the waveform, 
while the Q$_{part}$ denotes the integral of the tail of the waveform.}
\end{figure}

The delays due to the experimental setup, $\Delta t_{\textrm{delay}}^{\textrm{exp}}$,
between the time of the interaction and the time
at which the signal is recorded
were due to the response time of the liquid scintillator of
the Gd--LS detector,
$\Delta t_{\textrm{rsp}}$, and the delay, $\Delta t_{\textrm{cab}}$, 
in the cables. 
The value of $\Delta t_{\textrm{rsp}}$ for this type of Gd--LS detector 
was previously determined to be $\approx$\,30\,ns, 
see Fig.~\ref{fig:pulse shape}. 
The delay in the cables was estimated to 
be $\Delta t_{\textrm{cab}} \approx 14$\,ns 
and the time resolution of the ADC was 4\,ns.
Thus, $\Delta t_{\textrm{delay}}^{\textrm{exp}}$\,=\,48\,ns was estimated. 
% This estimate will be confirmed later. No forward references,
% especially without a chapter. 

Accordingly, the search window for prompt signals was opened at a
time difference, $\Delta t_{\textrm{sig-}\upmu} = 16 + \Delta t_{\textrm{delay}}^{\textrm{exp}} = 64$\,ns, 
between the muon tag and the first signal in the Gd--LS detector 
with E\,$\geq$\,0.5\,MeV$_{ee}$.
The search window was closed at $\Delta t_{\textrm{sig-}\upmu} = 348$\,ns.
This maintained an efficiency of almost 100\,\%.
If a second muon tag was found before the end of the window
or a muon was identified in the left-side lead wall, 
the window was closed already at the respective time.
As the total rate of identified muons in the right-side and left-side walls
was only about 1\,Hz, the probability for an early closure was
below $4 \times 10^{-7}$ and thus the efficiency was not affected.

A total of 15,494 prompt signals in the Gd--LS detector were selected
as the candidates for nuclear recoils.

\subsection{Discrimination against signals from electron recoils}
\label{sec:nuclear recoil selection}

The sample of prompt signals still contains a considerable background 
from electron recoils due to $\upgamma$, electron or muon interactions. 
The waveforms of the events in the Gd--LS detector can be used
to distinguish between nuclear and electron recoil events.
Figure~\ref{fig:pulse shape} shows the average waveforms for nuclear 
and electron recoil events as obtained for an identical detector
with an AmBe calibration measurement~\cite{Qiang2017}.  
The tails of the waveforms are different.
The parameter Q$_{total}$ denotes the total integral of the 
waveform from 0 to 208\,ns  while Q$_{part}$ denotes the integral 
from 76\,ns to 208\,ns. The observable $Dis$ 
\begin{equation}
Dis = \frac{Q_{part_{1}}+Q_{part_{2}}}{Q_{total_{1}}+Q_{total_{2}}} ~,
\end{equation}
where `1' and `2' stands for the two photo-multipliers of the Gd--LS
detector, was constructed to distinguish between prompt signals 
from nuclear recoils and electron recoils. 
%The $Dis$ selection will be described further in Section~\ref{sec:final nuclear recoil}.
%However, this discriminator alone was not enough
%to select neutron-induced events to a sufficient degree.

\subsection{Delayed signal from $\upgamma$ emission after neutron capture in the Gd doped liquid scintillator}
\label{sec:delayed gamma}

A coincidence between a prompt signal and a delayed
signal, presumably indicating a neutron capture, was also required 
to identify neutron-induced events.
After thermalization, neutrons are captured in the Gd--LS 
after a diffusion time, $\Delta t_{\upgamma\textrm{-}n}$. 
The diffusion time predominantly depends on the fraction of Gadolinium 
in the liquid scintillator, which is 0.5\% for the liquid scintillator EJ-335
used in this experiment. 
The  average value  of $\Delta t_{\upgamma\textrm{-}n}$ of $ \approx 7$\,$\upmu$s 
was determined with an AmBe neutron source, following the
procedure developed for an identical Gd--LS detector~\cite{Qiang2017}. 

\begin{figure}[t]
\centering
\includegraphics[width=0.48\textwidth]{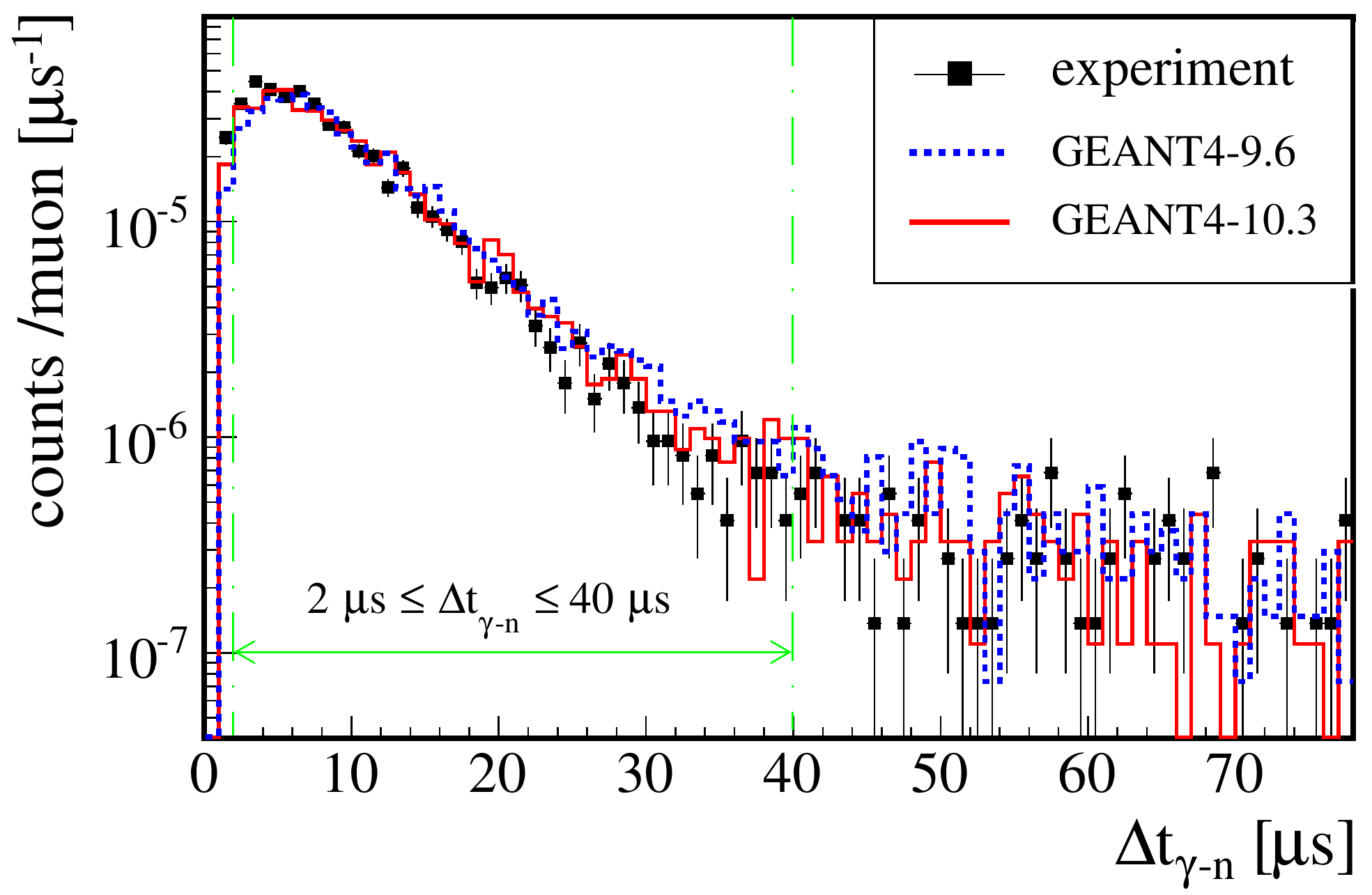}
\caption{\label{fig:diffuse time} The measured distribution of the 
time difference, $\Delta t_{\upgamma\textrm{-}n}$, between candidate signals 
for $\upgamma$ rays emitted after neutron capture and the prompt signals 
identified as nuclear recoils through the cut $Dis>0.12$. 
Also shown are two GEANT4 simulations, see Section~\ref{sec:simulation}, 
normalized to the integral of the measured distribution for  
$2\,\upmu$s$ < \Delta t_{\upgamma\textrm{-}n} < 40\,\upmu$s.}
\end{figure}

To study the coincidences as observed in the data taken in the underground 
laboratory in T\"ubingen, preliminary cuts were used to select
the nuclear recoils out of the prompt signals.
A cut of $Dis > 0.12$ was introduced and the energy of the prompt signal
was limited to 20\,MeV$_{ee}$.
The distribution of the difference between the time of the 
nuclear recoil and the time of the delayed signal,
$\Delta t_{\upgamma\textrm{-}n}$, is shown 
in Fig.~\ref{fig:diffuse time}. Also shown are the predictions 
from two GEANT4 simulations, see Section~\ref{sec:simulation}.
The predictions are normalized to the data in the time interval
from 2 to 40\,$\upmu$s. 
The shape of the measured distribution
is quite well described. This shows that 
the preliminary selection of nuclear recoils is quite effective 
and that the predictions from simulation can be used to
evaluate the efficiency of a $\Delta t_{\upgamma\textrm{-}n}$ cut.

\begin{figure}
\centering
\includegraphics[width=0.48\textwidth]{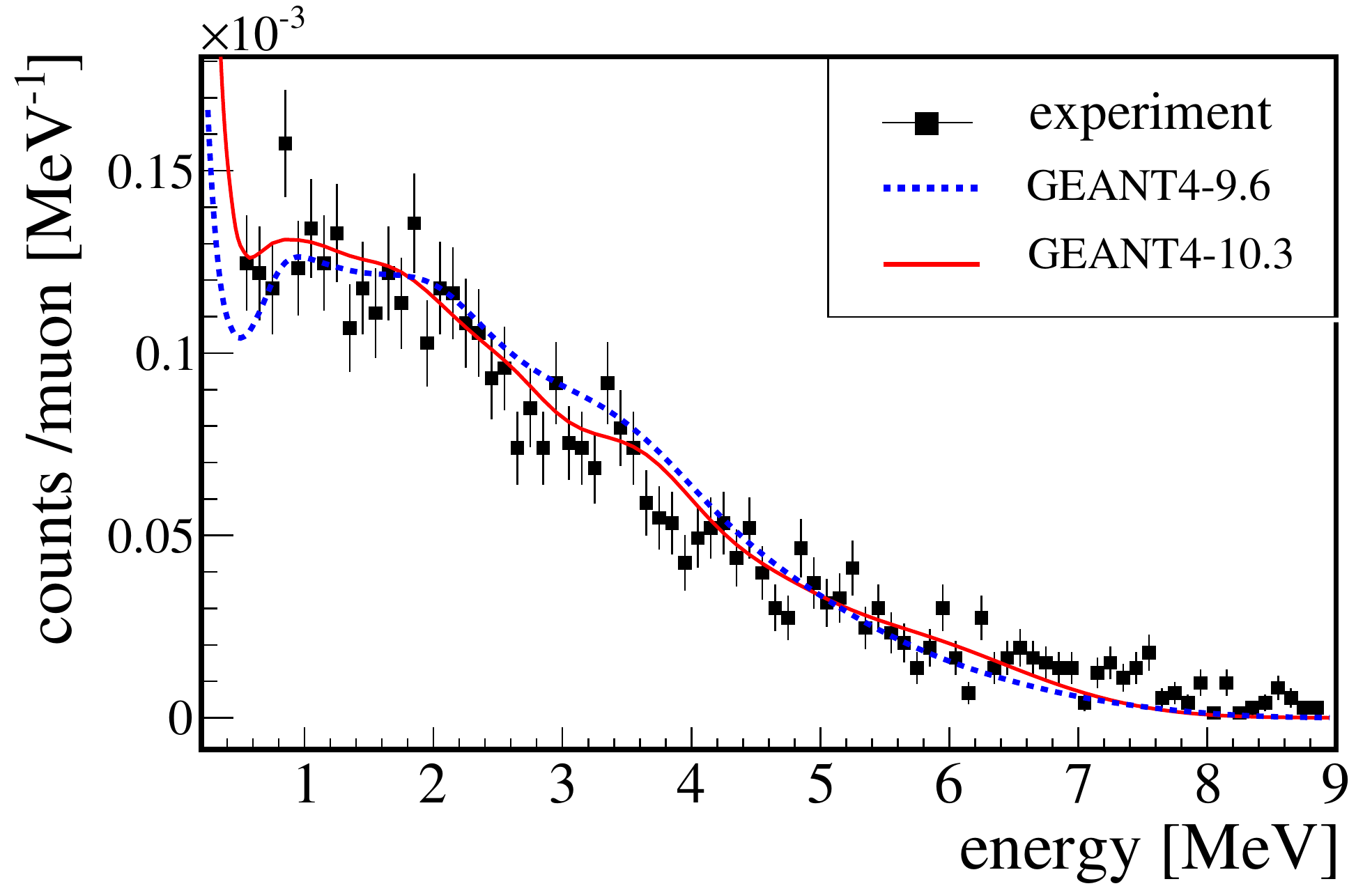}
\caption{\label{fig:delayed spectrum} The distribution of the energy 
observed for time-selected delayed signals
following a prompt signal identified as a nuclear recoil by the cut $Dis>0.12$.
Also shown are predictions from two GEANT4 simulations which
are normalized to the integral of the measured spectrum in the interval from 
0.5\,MeV to 9\,MeV.}
\end{figure}

The time window of $2\,\upmu$s$ < \Delta t_{\upgamma\textrm{-}n} < 40\,\upmu$s
was chosen to select delayed $\upgamma$ signals from neutron capture.
This rejected most of the accidental background 
while, according to simulation, retained an efficiency
of $\approx\,84$\,\%. The distribution of the energy deposited by the time-selected
candidates for delayed $\upgamma$ signals is shown 
in Fig.~\ref{fig:delayed spectrum}. 
Again, the predictions from the GEANT4 simulations describe the data
quite well.
The $\Delta t_{\upgamma\textrm{-}n}$ selection was combined with the
requirement that, 
the energy of the delayed signal was in the range of 
0.5 to 9\,MeV.

\subsection{Final selection of nuclear recoil events}
\label{sec:final nuclear recoil}

\begin{figure}
\centering
\includegraphics[width=0.48\textwidth]{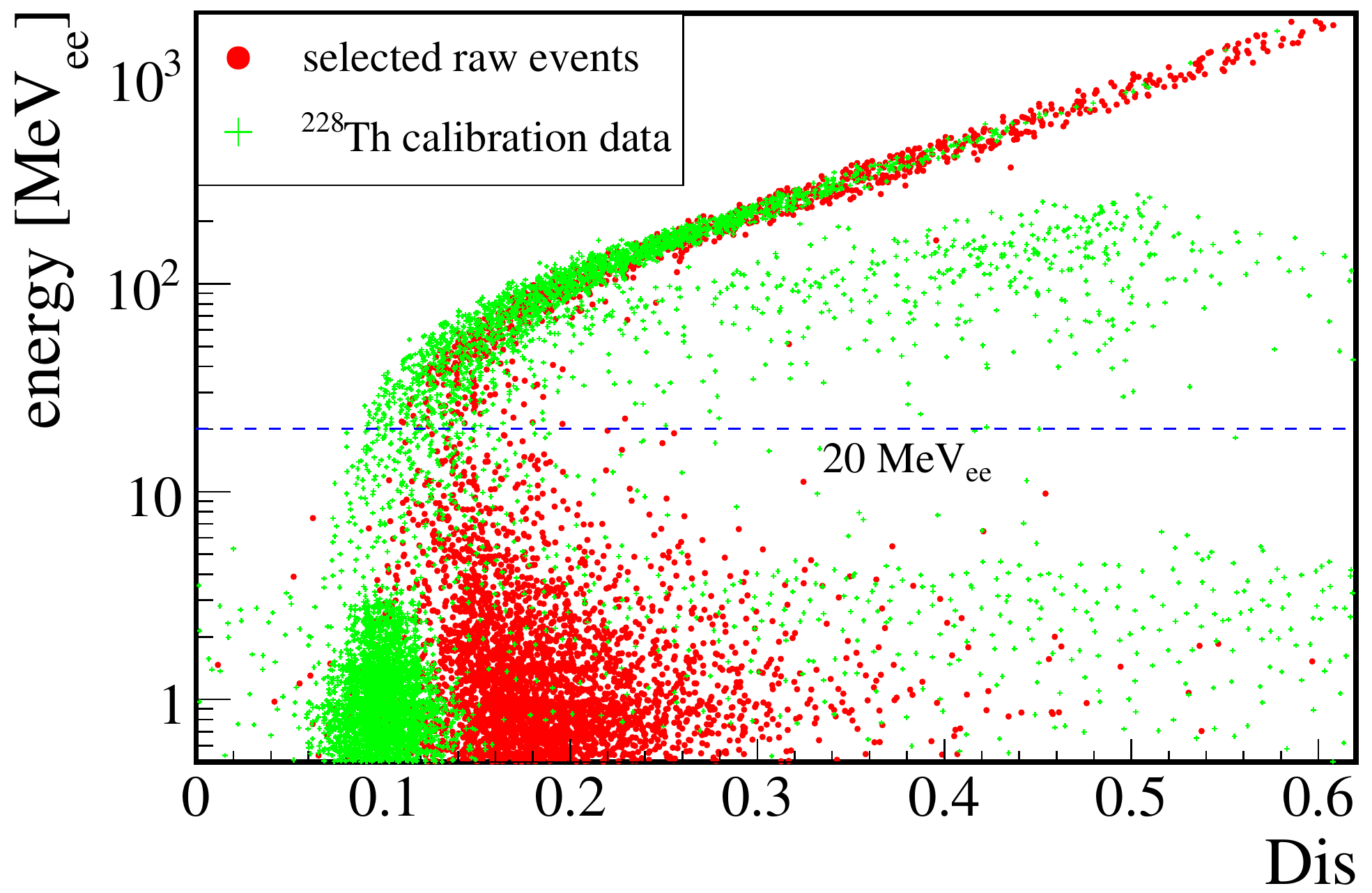}
\caption{\label{fig:scatters of nuclear recoils} Energy versus $Dis$ 
discriminator for the prompt signals of time-selected events 
which are candidates for nuclear recoil signals.
Also shown are events recorded for calibration data recorded with 
a $^{228}$Th $\upgamma$-source. The cut valued for 20\,MeV$_{ee}$ is indicated
by a horizontal dashed line.} 
\end{figure}

Without the requirement that the energy recorded for the prompt signal
was below 20\,MeV$_{ee}$ and  preliminary $Dis>0.12$ cut, 
4296 events within the time window from 2 to 40\,$\upmu$s 
were selected. 
Figure~\ref{fig:scatters of nuclear recoils} shows a scatter plot 
of the energy of the prompt signals versus $Dis$ for these events.
Also shown is the distribution of events from a calibration measurement with
a $^{228}$Th $\upgamma$-source.
The two distributions are not normalized to each other in any way.
The events from the calibration data shown in 
Fig.~\ref{fig:scatters of nuclear recoils}
with an energy below 4\,MeV$_{ee}$ are predominantly due to $\upgamma$ rays 
from the $^{228}$Th source and form a clearly different class of events.
However, some events populate the nuclear recoil region. They are mainly due 
to environmental radioactivity.
The events above 4\,MeV$_{ee}$ 
are due to cosmic radiation including muon-induced neutrons. 

Below 20\,MeV$_{ee}$, the two event classes are clearly separated.
Above 20\,MeV$_{ee}$, nuclear and electron recoils can not be 
distinguished efficiently because of changes in the waveforms.
Thus, the requirement that the energy of the prompt signal was 
below 20\,MeV$_{ee}$ was retained. This reduces the sample by about
300 events.

\begin{figure*}
\centering
\begin{subfigure}{0.48\textwidth}
\includegraphics[width=\linewidth]{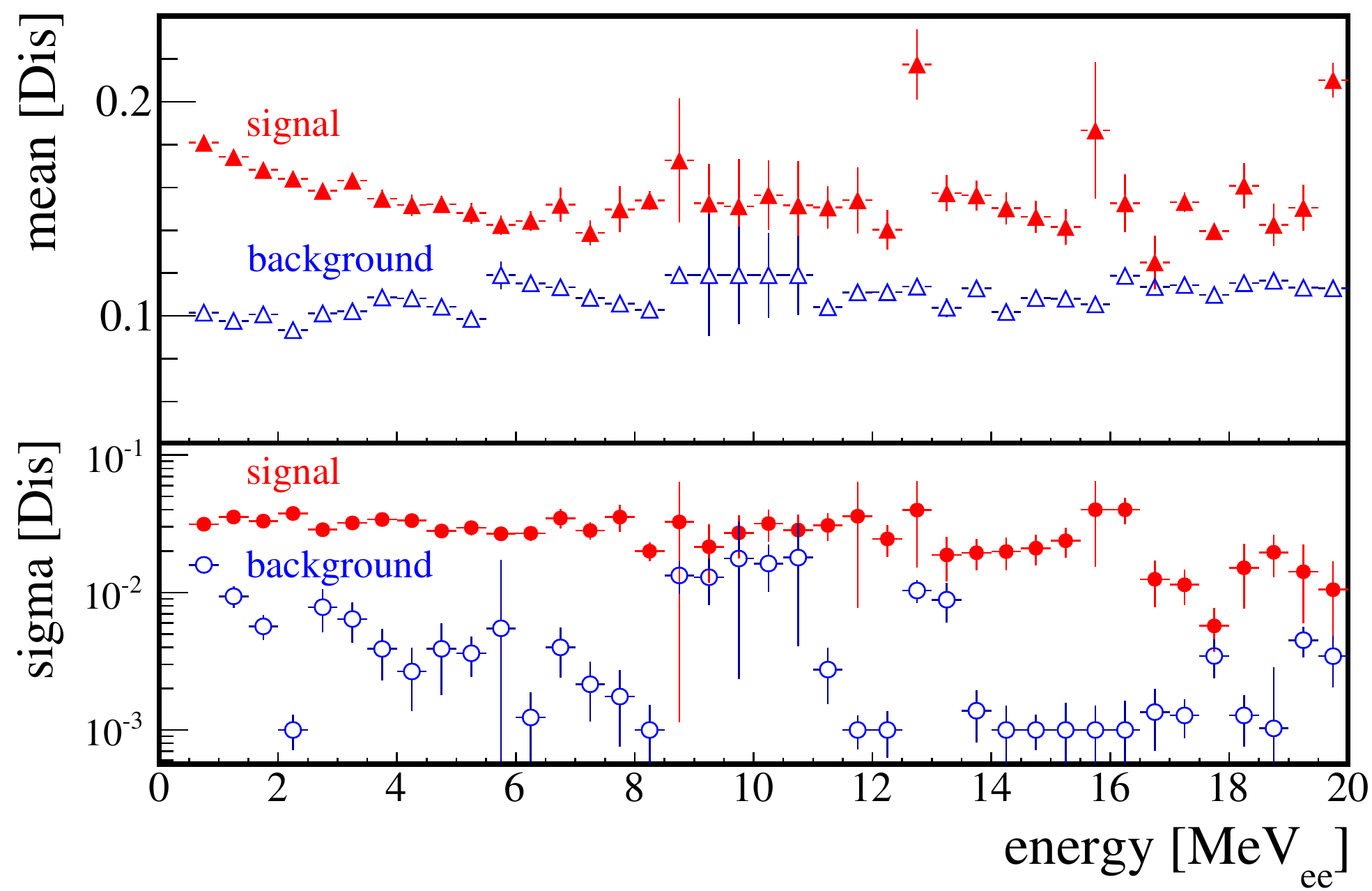}
\caption{\label{fig:mean and sigma} }
\end{subfigure}
\hfill
\begin{subfigure}{0.48\textwidth}
\includegraphics[width=\linewidth]{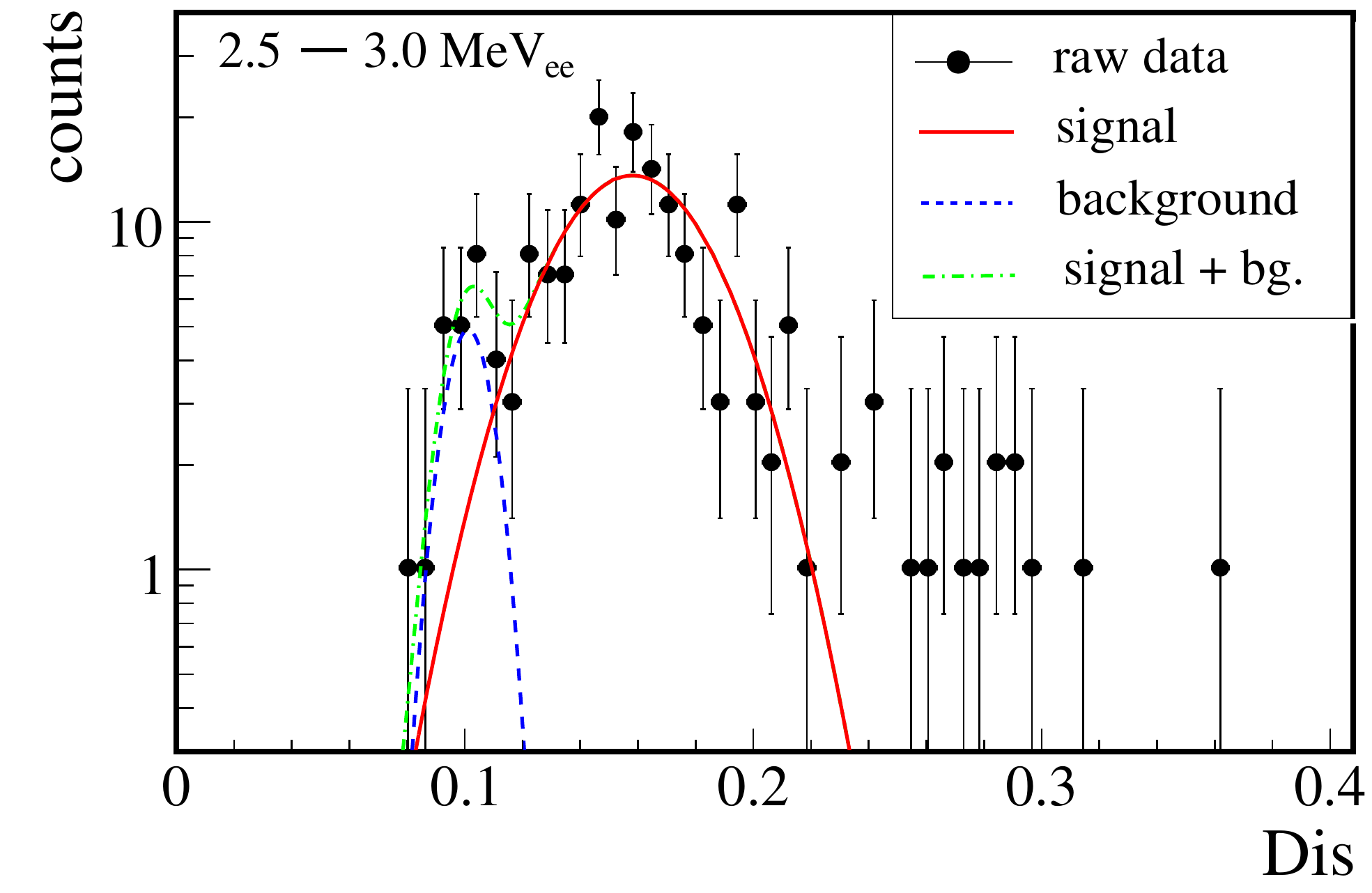}
\caption{\label{fig:2.5-3.0MeV} }
\end{subfigure}
\caption{\label{fig:double fit} Unbinned Maximum-Likelihood fit of two Gaussian 
functions to the $Dis$ distributions. (a) The mean (top panel) 
and sigma (bottom panel) values of the Gaussians from the fits, 
both for signal and background; 
(b) Results of the Gaussian fit for the energy bin 2.5\,--\,3.0\,MeV$_{ee}$;
    the $\chi^2$/dof is 1.08 for this fit.}
\end{figure*}

For the final determination of the number of events, 
the usage of the $Dis$ discriminator was refined.
It was used in an energy-dependent way in order
to discriminate as effectively as possible
against electron recoils in the time-selected events
with an energy of the prompt signal below 20\,MeV$_{ee}$.  
An unbinned Maximum-Likelihood method was used to fit $Dis$ with 
two Gaussian functions for each energy bin of 0.5\,MeV$_{ee}$ width.
While the integral of the Gaussian with lower mean $Dis$ 
is taken as the number of background events, 
the integral with the higher mean $Dis$ is taken as the number of 
events attributed to nuclear recoils.
Figure~\ref{fig:mean and sigma} shows the mean and sigma values of 
the Gaussians from the fits. 
For some of the bins, the mean and sigma values are fluctuating due to 
a lack of statistics. This is taken into account through the uncertainties 
as given by the fits. 
Figure~\ref{fig:2.5-3.0MeV} depicts the fit results on the discriminator 
$Dis$ for the typical energy bin from 2.5\, to \,3.0\,MeV$_{ee}$. 
The good quality of the fit indicates that very few events with high
$Dis$ are lost.

In summary, nuclear recoil events were selected which comprised
\begin{itemize}
\item a muon tag indicating the passage of a muon through the 
      right-side lead wall,
\item followed by a prompt signal in the Gd--LS detector with
      an energy in the range of 0.5 to 20\,MeV$_{ee}$ and
      a time delay 64\,ns $ < \Delta t_{\textrm{sig-} \upmu} < 348$\,ns,
\item followed by a delayed signal in the Gd--LS detector with
      an energy in the range of 0.5 to 9\,MeV and
      a time delay $2\,\upmu$\,s $ < \Delta t_{\upgamma\textrm{-}n} < 40\,\upmu$s.   
\end{itemize}
       
The final number of events was determined using fits to the $Dis$
discriminator on the prompt signals in individual energy bins. 
The fits determined the number of background events to be 457$\pm$35. 
The overall number of signal events was determined to be
3534$\pm$68. 
The event numbers of nuclear recoils as determined from the fits were used for 
further analysis.

\subsection{Systematic uncertainties on the selection of nuclear recoil events}

The systematic uncertainties had two main sources:
\begin{itemize}
\item changes in the energy scale;
\item uncertainties due to the time resolution of the ADC.
\end{itemize}

The energy scale as determined by the energy calibration,
see Section~\ref{sec:energy calibration},
was not completely stable during the 5 months of data taking.
The light yield of the liquid scintillator varied without a trend
by 4.6\,\%.
Folding in the energy resolution, 
the uncertainty on the energy scale was conservatively estimated 
to be $\approx\,10$\%. 
When varying the offline threshold $E_{thr}=0.5\,$MeV by 10\,\%, 
the number of signal events changes by 8.1\,\%. 

The time resolution of the ADC affects the $\Delta t_{\textrm{sig-}\upmu}$ cut 
selecting prompt signals, see Section~\ref{sec:nuclear recoil}.
When varying the $\Delta t_{\textrm{sig-}\upmu}$ cut by 4\,ns, 
the number of signal events changes 7.5\%. All other systematic uncertainties are negligible. 
The signal for nuclear recoil events with a delayed $\upgamma$ signal
for a lifetime of 151.6 days 
becomes $3534\pm68\,(\textrm{stat.})\pm390\,(\textrm{syst.})$.  

\subsection{Selection of capture events}

The requirement of an observable nuclear recoil excludes neutrons 
with low energies.
Such neutrons are, however, abundantly produced in muon-induced interactions
and they can be captured in the Gd--LS detector.
The selected events comprise a muon tag and a delayed signal:

\begin{itemize}
\item a muon tag indicating the passage of a muon through the 
      right-side lead wall,
\item followed by a delayed signal in the Gd--LS detector with
      an energy in the range of $E_{thr}^{\textrm{CE}} = 2.0$ to 9\,MeV and
      a time delay $2\,\upmu$s $ < \Delta t_{\upgamma\textrm{-}\upmu} < 40\,\upmu$s.   
\end{itemize}

\begin{figure}
\centering
\includegraphics[width=0.48\textwidth]{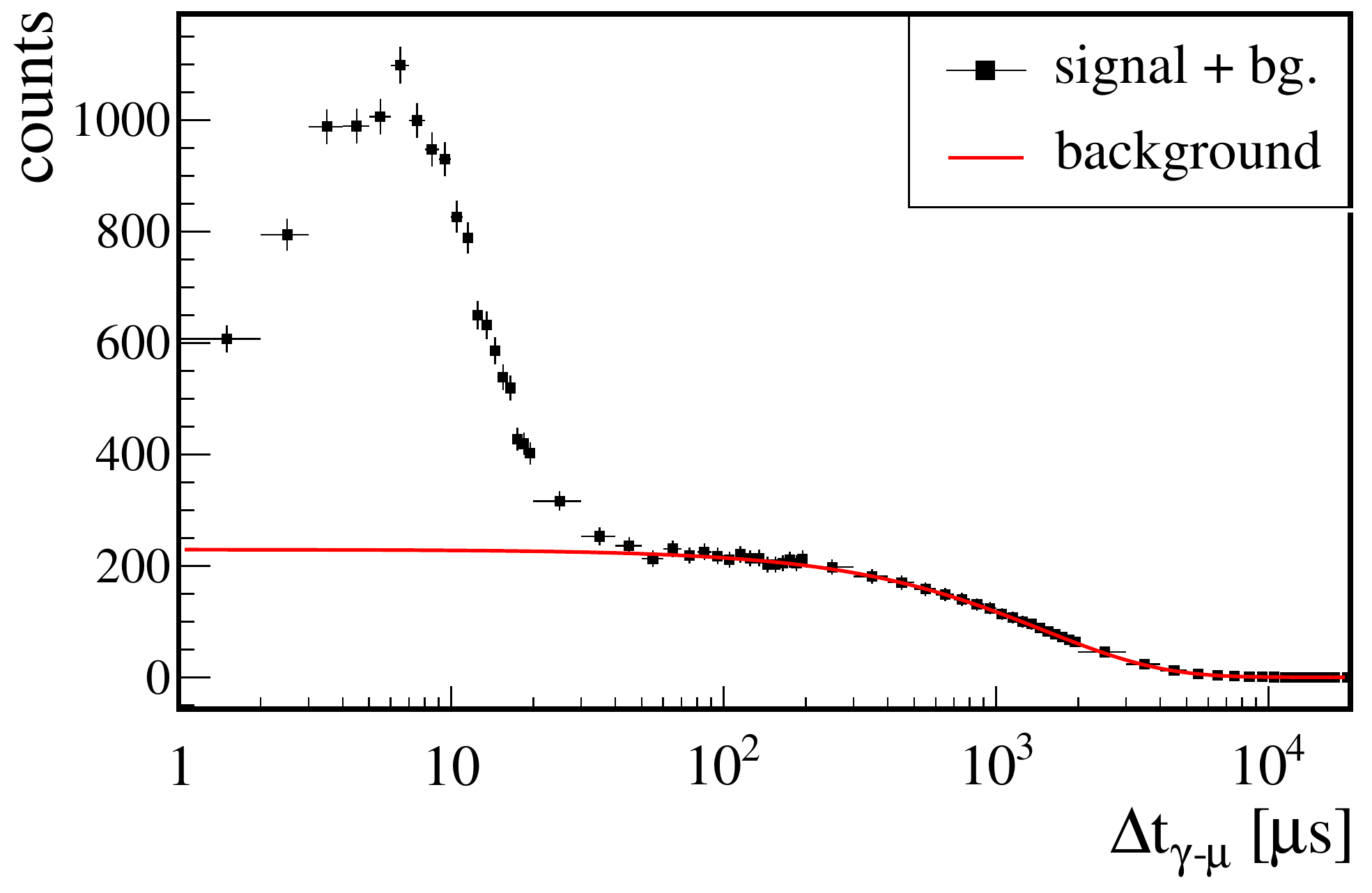}
\caption{\label{fig:capture time} The distribution of the time delay,
$\Delta t_{\upgamma\textrm{-}\upmu}$,  
of the neutron capture signals with $E_{thr}^{\textrm{CE}}$\,=\,2\,MeV 
with respect to the muon tags.
The result of a fit of an exponential to the background for 
$\Delta t_{\upgamma\textrm{-}\upmu} > 40\,\upmu$s is shown as a solid line.}  
\end{figure}

\begin{figure}[t]
\centering
\includegraphics[width=0.48\textwidth]{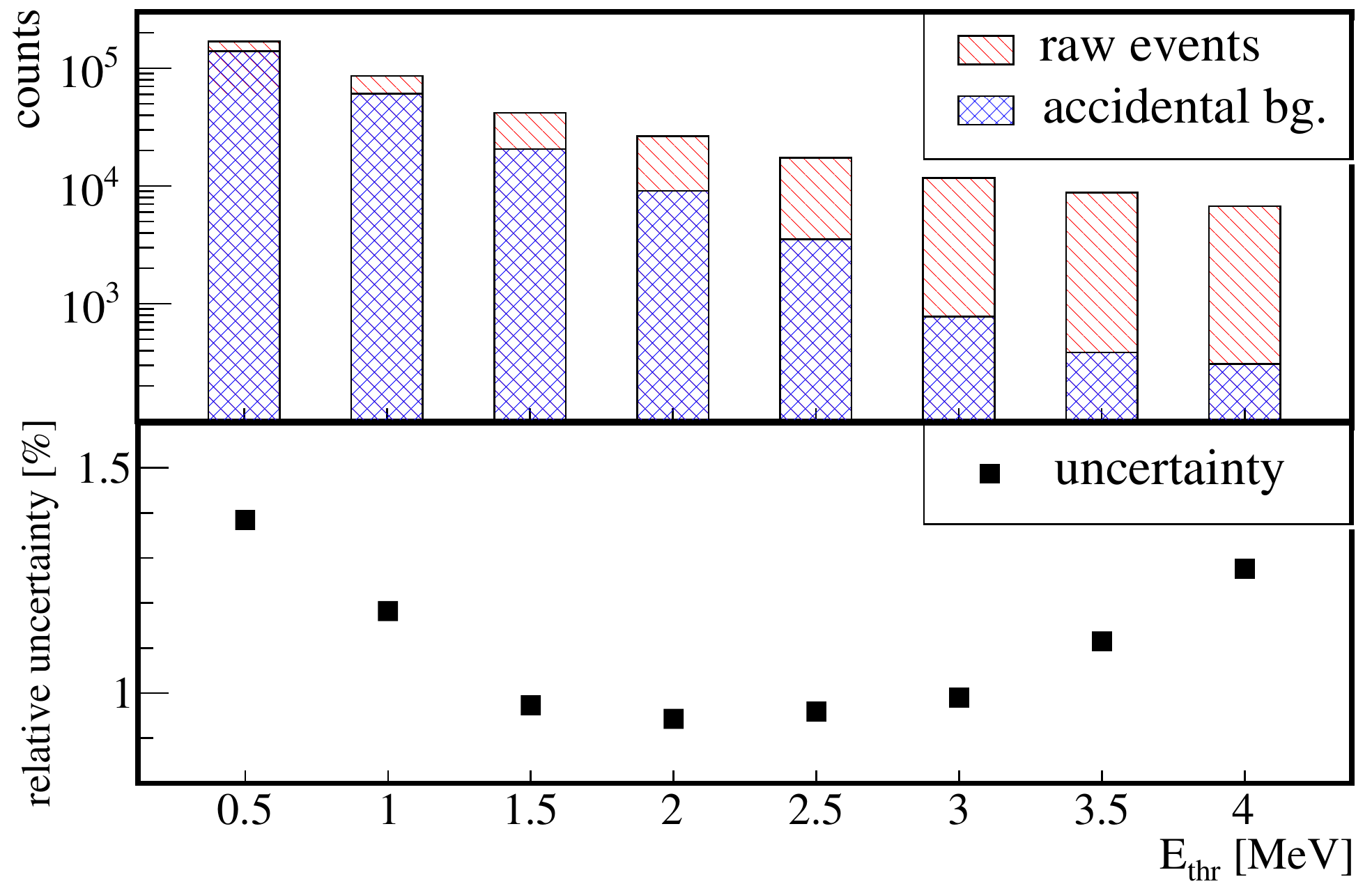}
\caption{\label{fig:capture uncertainty} Top panel: Raw number of selected 
capture events (raw events) and the contribution of accidental background (bg.)
events as determined by exponential fits for different  $E_{thr}^{\textrm{CE}}$.
Bottom panel: Relative statistic uncertainties on the background subtracted
event numbers.}
\end{figure}

The observed $\Delta t_{\upgamma\textrm{-}\upmu}$ distribution is shown in Fig.~\ref{fig:capture time}.
%%%%%The distribution is normalized to the number of muon tags.
Without the requirement of a coincident nuclear recoil, the background
level is substantial. However, Fig.~\ref{fig:capture time} demonstrates
that signal events do not occur beyond  
$\Delta t_{\upgamma\textrm{-}\upmu} = 40\,\upmu$s. The accidental 
background was determined by fitting an exponential to the  
$\Delta t_{\upgamma\textrm{-}\upmu} > 40\,\upmu$s distribution. It was used
to subtract the accidental background statistically.

The value of $E_{thr}^{\textrm{CE}} = 2.0$\,MeV was chosen to minimize the
statistical uncertainty on the background subtracted event number.
Figure~\ref{fig:capture uncertainty} shows the raw number 
of selected neutron capture events  and the number of 
accidental background events according to the fits as well as the
resulting relative uncertainties depending on $E_{thr}^{\textrm{CE}}$.
The uncertainty has a shallow, but distinct minimum at 
 $E_{thr}^{\textrm{CE}} = 2.0$\,MeV.

For  $E_{thr}^{\textrm{CE}} = 2.0$\,MeV, a total of $ 26399 \pm 162 $ events 
were selected including  $ 9082 \pm 21 $ background events.

\section{Monte Carlo}
\label{sec:simulation}

The MC simulation was performed in two stages:
\begin{enumerate}
\item propagation of muons from the surface to the underground lab;
\item muon interactions and physics in the experiment.  
\end{enumerate}

\begin{figure}
\centering
\includegraphics[width=0.48\textwidth]{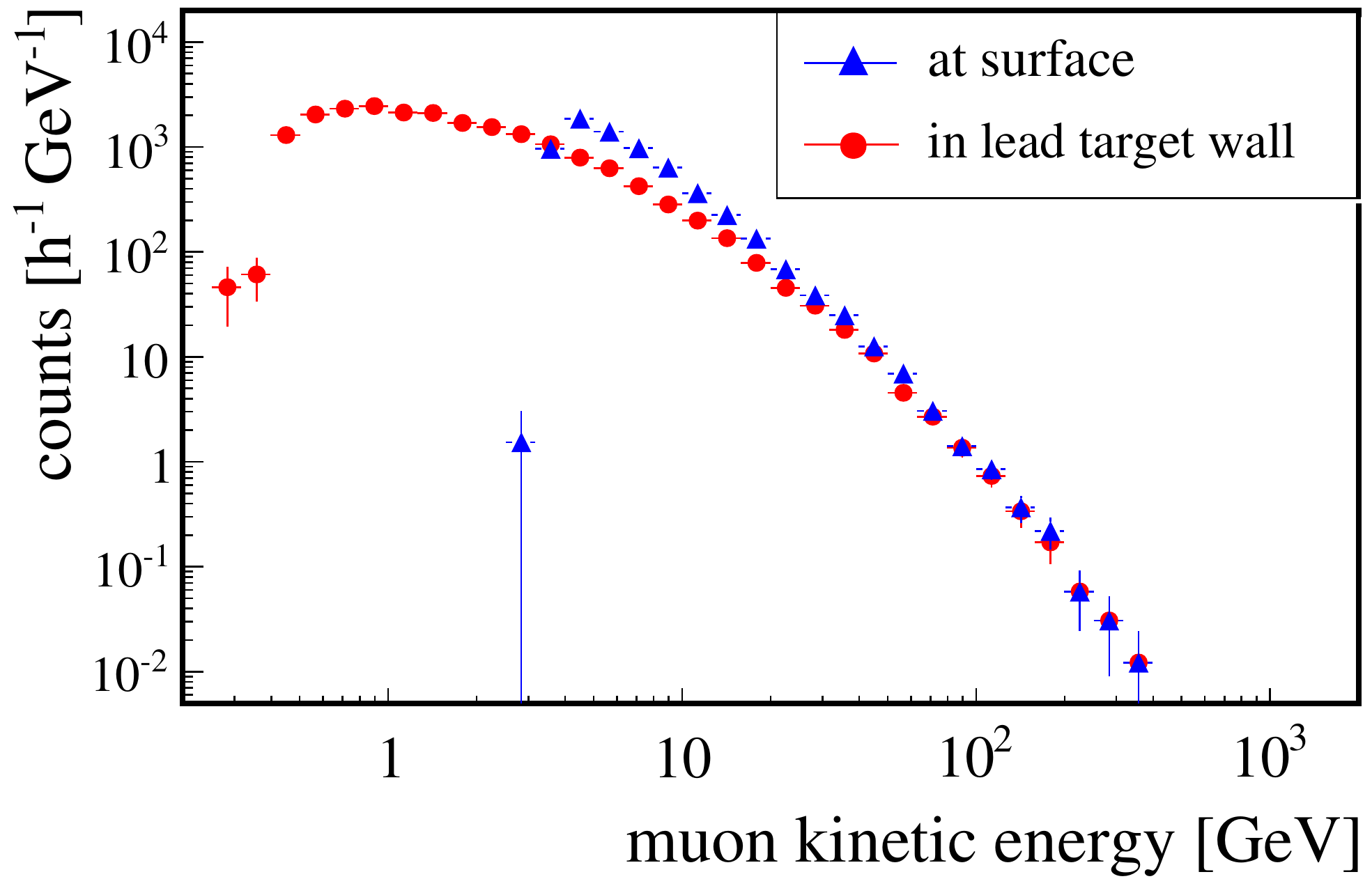}
\caption{\label{fig:mulab} Distribution of the kinetic energy of tagged muons
  at the surface and inside the lead wall simulated by FLUKA version 2011.2c. The latter was computed as the
  average of the kinetic energies above and below the wall.
}
\end{figure}

For the first stage, FLUKA-2011.2c~\cite{fluka-1,fluka-2} was used to simulate 
the propagation of cosmic muons through the overburden 
into the underground laboratory.
The geometry was implemented according to the technical drawings where the
material of the overburden was described as ``densified soil'', for which
a density of 2.25\,g/cm$^3$ was assumed.
The average energy of the 
muons entering the laboratory from the top was found to be $\approx 7$\,GeV.
The distributions of the kinetic energy of tagged muons at the surface
and inside the lead wall are shown in Fig.~\ref{fig:mulab}.
The average kinetic energy of tagged muons is 11.2\,GeV on the surface and 
$\approx 8$\,GeV inside the lead wall.

The energy and angular distributions of cosmic muons above ground were
parametrized~\cite{Bogdanova}. 
A comparison to the world data on the muon momentum spectrum at sea-level
available from the PDG~\cite{PDG-2014} showed a good agreement in the region of
interest.
The underground laboratory is located at 452 meter above sea-level which
causes a few percent increase of the muon flux as opposed to sea-level.
In addition, low energy muons are subject to solar cycle, weather
and seasonal atmospheric changes. 
Therefore, we assume a systematic uncertainty on the integral muon
intensity of the order of 15\%. 
However, the absolute flux is not important because all predictions are
normalized to the number of observed muon tags.  

A cross-check was performed starting the simulation from cosmic rays
interacting in the earth atmosphere. A good agreement on energy and
angular distributions in the laboratory was found.
The mean energy of muons in the lab was confirmed within 5\%.
The angular distributions were also similar. The important quantity is
the average path-length of 51\,cm for tagged muons, which was
confirmed within 2\%.

The positions and momenta 
of the muons and their secondary particles  
on the surface of a virtual sphere with 1\,m radius around 
the combined setup of MINIDEX and the Gd--LS detector
were saved together 
with information on the relative time of arrival.
This was the input to stage~2 of the simulation.

For the second stage, GEANT4 was used to simulate: 
\begin{itemize}
\item the propagation of all muons and muon-induced secondary particles 
      from the virtual sphere towards MINIDEX, 
\item the production of muon-induced neutrons and the production
      of all other associated particles, 
\item the propagation of the neutrons and all other particles 
      to the Gd--LS detector, 
\item the response of the Gd--LS detector. 
\end{itemize}

The optical photon processes, i.e.\
the generation of scintillation light, light transportation and
light collection, were not simulated.
The energy depositions were taken as simulated by GEANT4.
For nuclear recoils, the reduction of the light output due
to quenching~\cite{Birks} was corrected for according to
previous measurements for the Gd--LS type
EJ-335~\cite{Qiang-Quenching,Qiang2017}. 

Two versions of GEANT4 were used:
\begin{itemize}
\item GEANT4--9.6.p04 with the Shielding~2.0 modular physics list 
      with the muon-nuclear reaction switched on by hand, 
    \item GEANT4--10.3.p02 with the ShieldingM~2.1 modular physics list
      for which the muon-nuclear reaction is switched on by default.
\end{itemize}

The Shielding physics list was introduced for GEANT4-9.4.
It was originally developed for studies related to ion-ion collisions  
and to the penetration power of neutrons.
It is also suited to study aspects of neutron physics for 
underground and low background experiments. 
The Shielding physics list~\cite{Shielding1} contains a 
well motivated selection of simulation packages
for electromagnetic and hadronic physics processes.
Its high energy part is taken from the FTFP\_BERT~\cite{BERT_HP} physics list. 
It uses FTFP and Bertini to simulate proton, neutron, pion and kaon interactions. 
For neutrons with an energy below 20\,MeV, the list neutron\_HP~\cite{BERT_HP}
is used. 
The QMD model~\cite{Shielding2} is used to simulate the interactions of ions. 

The ShieldingM physics list~\cite{ShieldingM}, 
which was developed for muon experiments,
was introduced since GEANT4-10.1.  
It is based on the Shielding physics list with transition from
the Bertini to the FTFP model happening in the region
between 9.5 and 9.9\,GeV.

The production threshold for protons in GEANT4-9.6 is by default 70\,keV
while in GEANT4-10.3 it is 0\,keV to allow for low energy nuclear recoils.
The threshold for secondary particle production is implemented as a
distance cut in GEANT4. Specifically for this study, it
was set to 0.1\,mm in both GEANT4 versions. 
This corresponds to energy thresholds of $\approx 30$\,keV,
$\approx 240$\,keV and $\approx 230$\,keV 
for $\upgamma$, e$^{-}$ and e$^{+}$ production in lead, 
and $\approx 1$\,keV and $\approx 80$\,keV for $\upgamma$ and 
e$^{-}$/e$^{+}$ production in the Gd--LS. 
These values are well below the thresholds for $\upgamma$ and e$^{-}$/e$^{+}$
nuclear reactions, and well below the Gd--LS detector energy threshold, 
see Section~\ref{sec:energy calibration}.

The dashed curve in Fig.~\ref{fig:total spectrum} represents the 
spectrum of the energy deposited by cosmic muons and their secondary particles 
in the Gd--LS detector as predicted by GEANT4-9.6.
The data are reasonably well described above an energy of 4\,MeV, especially in 
the region of minimum ionizing particles. The discrepancy of the tail towards very high energies
is due to saturation of the PMT readout and not modeled in simulation.
This has, however, no influence
on the experiment. Below 4\,MeV, the data include events from
natural radioactivity which was not part of the simulation.

The dashed curve in Fig.~\ref{fig:muon spectrum} represents the
prediction for the spectrum of the energy deposited in 
muon scintillator panel~1 for events with a muon tag.
The position of the peak and the tail towards higher energies are well
described. The tail in the data toward lower energies is not described
as no inefficiencies were simulated.

\begin{figure*}
\centering
\begin{subfigure}{0.48\textwidth}
\includegraphics[width=\linewidth]{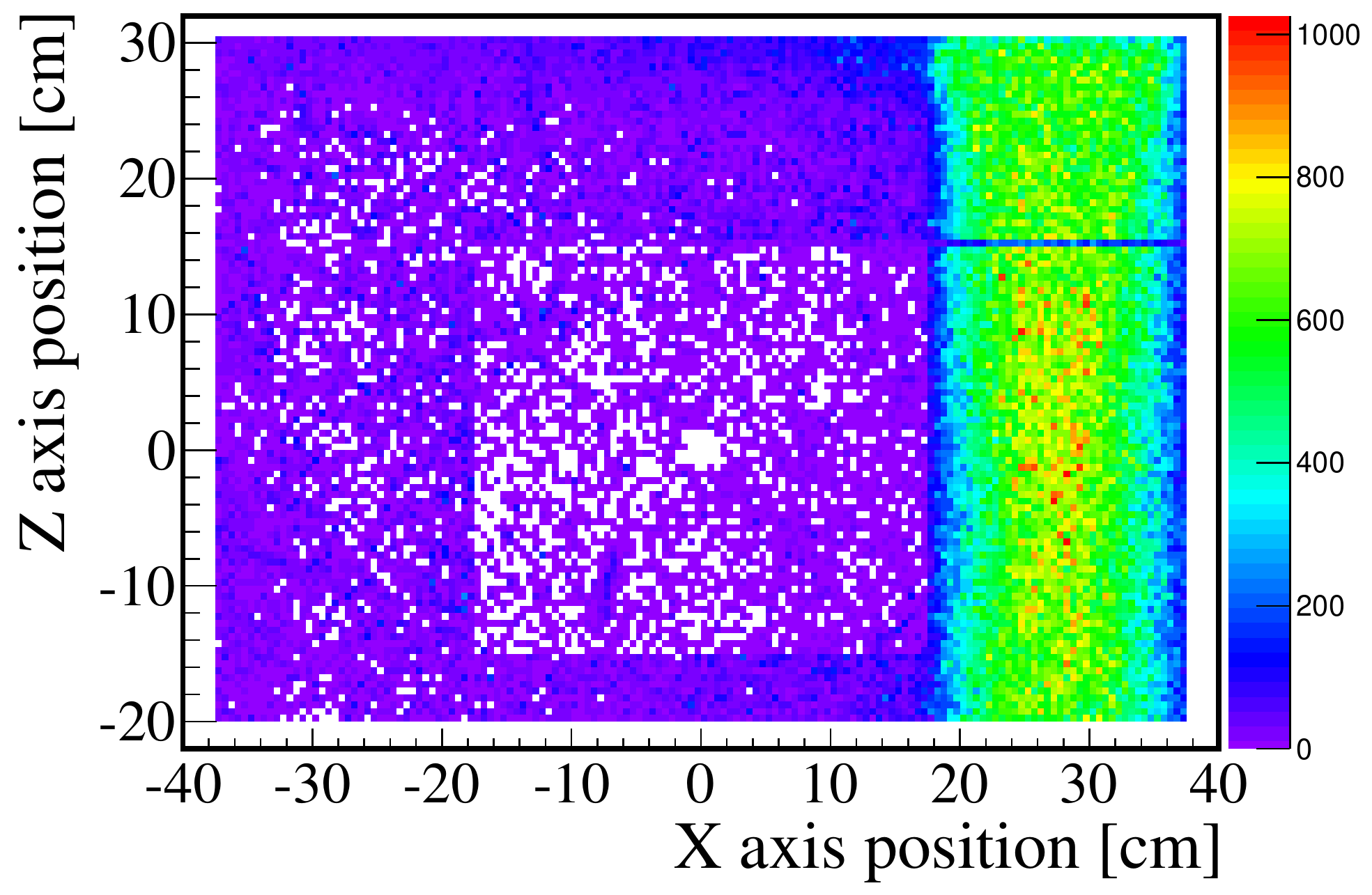}
\caption{\label{fig:position distribution}  }
\end{subfigure}
\begin{subfigure}{0.48\textwidth}
\includegraphics[width=\linewidth]{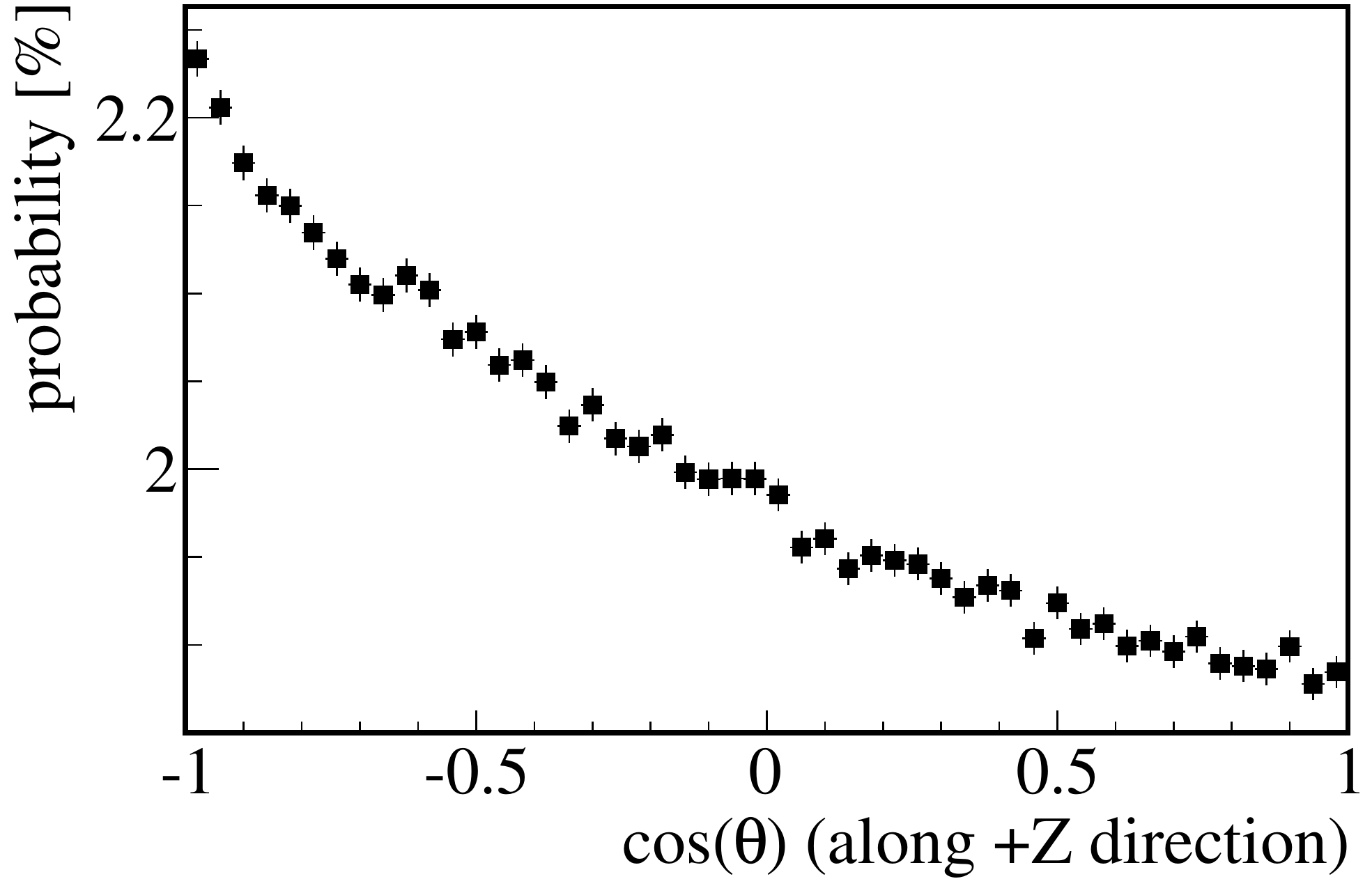}
\caption{\label{fig:angular distribution} } 
\end{subfigure}
\caption{\label{fig:simulated distribution} 
Predicted (GEANT4-9.6) distributions of (a) the production
position and (b) the azimuth angle of neutrons induced in lead
in events with a muon tag, 
see Fig.~\ref{fig:setup} for the definition of the X and Z axis.}
\end{figure*} 

The number of simulated muon tags was $\approx\,28.7\times 10^{6}$.
The average energy of the tagged muons was $\approx 8$\,GeV.
The GEANT4-9.6 predicted distributions of the position of neutrons produced
in events with a muon tag is shown in Fig.~\ref{fig:position distribution}. 
Here, only primary neutrons, i.e.\ neutrons produced directly by 
the tagged muons, and neutrons produced by muon-induced particles 
other than neutrons were considered.
Secondary neutrons generated by primary neutrons were not considered. 
The production distribution follows the structure of the MINIDEX setup, 
see Section~\ref{sec:setup}. Most of the neutrons are generated inside the 
right-side lead wall (17.5\,cm\,$<$\,X\,$<$\,37.5\,cm), which 
was used for tagging.
The neutrons produced outside the right-side lead wall   
were generated by muon-induced particles such as $\upgamma$, 
electrons, protons, pions and other particles. 

The angular distribution of the neutrons is shown 
in Fig.~\ref{fig:angular distribution}.
The distribution shows that neutrons are predominantly produced
downwards, but that this effect is not very strong.
The number of upward going neutrons is predicted to be
$\approx 85$\,\% of the number of downward going neutrons.

\begin{figure}[t]
\centering
\includegraphics[width=0.48\textwidth]{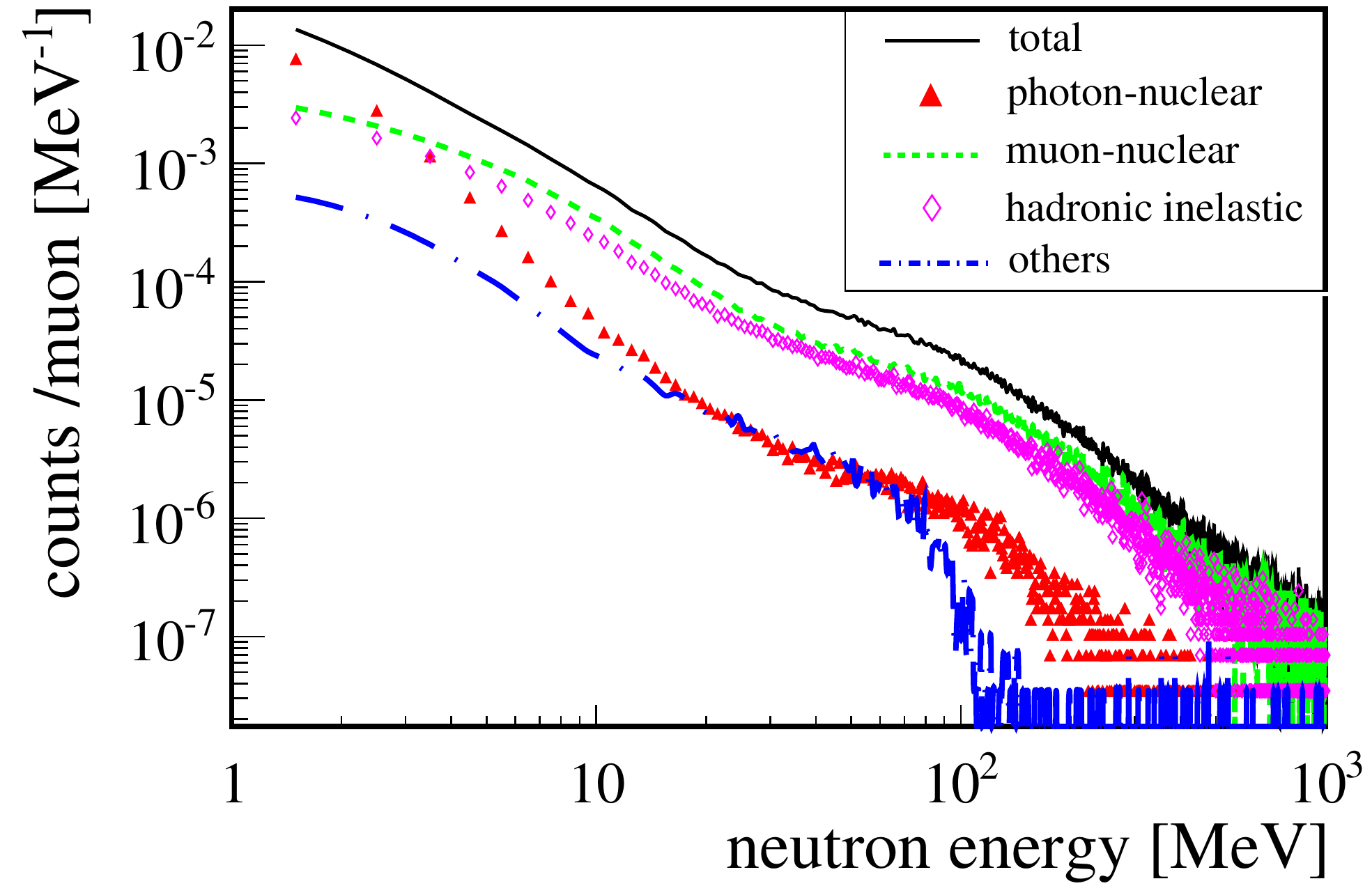}
\caption{\label{fig:simulated neutron energy spectrum} 
The predicted (GEANT4-9.6) energy spectrum of neutrons produced in 
events with a muon tag. Also shown are the individual contributions 
from photon-nuclear, muon-nuclear, and hadronic-inelastic interactions.
Other contributions include e$^{-}$/e$^{+}$ nuclear reactions as well as 
$\upmu^{-}$/$\uppi^{-}$/K$^{-}$ captures and decays.}
\end{figure}

\begin{figure}
\centering
\includegraphics[width=0.48\textwidth]{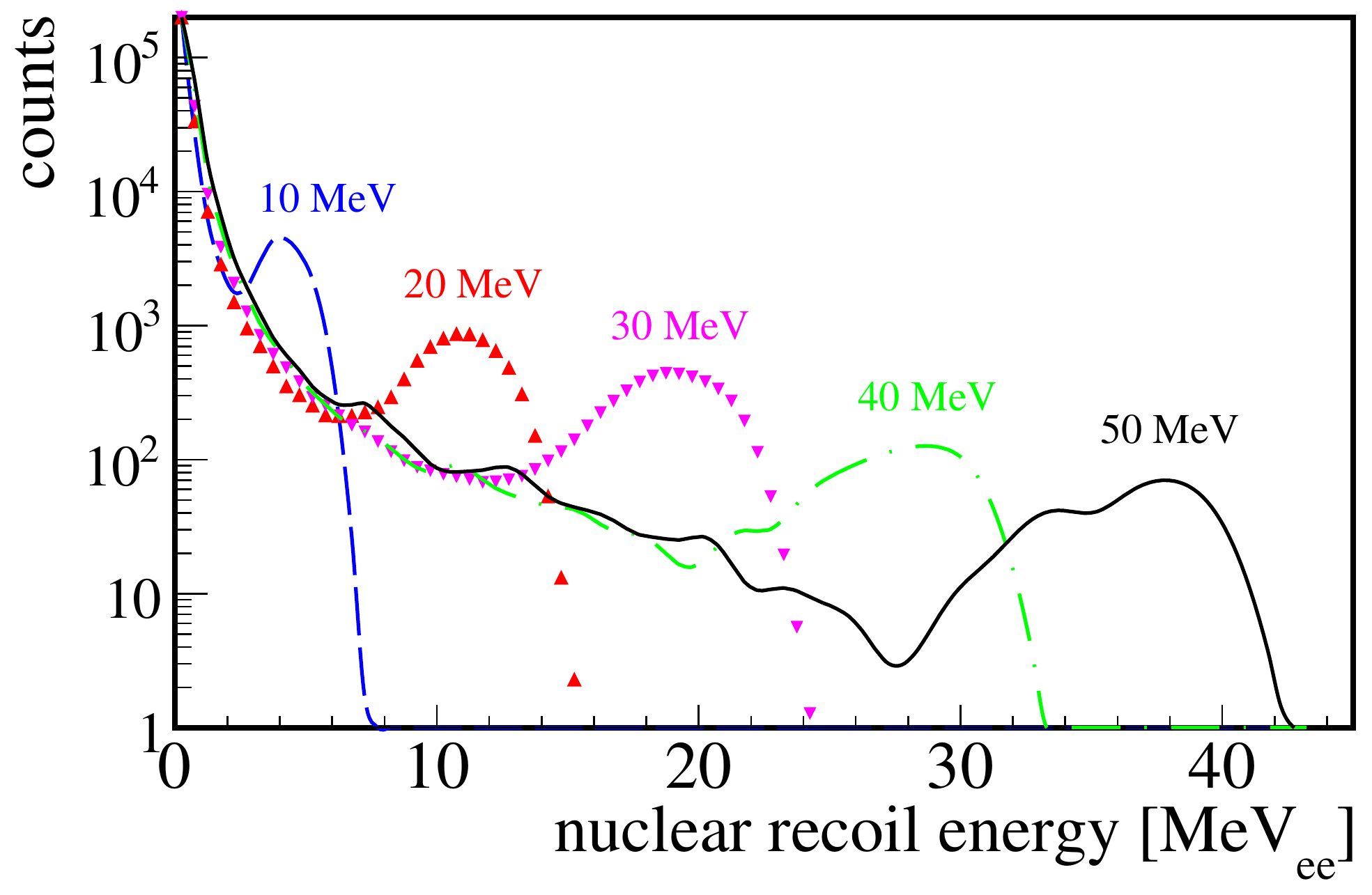}
\caption{\label{fig:response function} Simulated (GEANT4-9.6)
  response functions of the Gd--LS detector for 10, 20, 30, 40 and 50\,MeV
  mono-energetic neutrons homogeneously distributed in the lead.
  The signal quenching for nuclear recoils in the Gd-LS was included in
  the simulation.}
\end{figure}

The predicted spectrum of the neutrons is shown in
Fig.~\ref{fig:simulated neutron energy spectrum}. 
Also shown are the individual contributions from the most important
processes of neutron production. For energies above 2\,MeV,
the dominant process is indeed the production through muon-nuclear processes.
However, there is substantial contribution from hadronic inelastic scattering
where the hadrons are created in a muon-induced shower.

The GEANT4 was also used to simulate the response of the
Gd--LS detector to nuclear recoils. 
Figure~\ref{fig:response function} shows selected examples 
for mono-energetic neutrons with energies of 10, 20, 30, 40 and 50\,MeV and
production positions homogeneously distributed in the right-side lead wall
with an isotropic angular distribution. 

The time structure of the events as predicted by GEANT4 was already 
presented in
Figs.~\ref{fig:simulated time interval to muon trigger} and 
\ref{fig:diffuse time}. 
In Fig.~\ref{fig:diffuse time}, the predicted distribution of
$\Delta t_{\upgamma\textrm{-}n}$,
the time between the prompt nuclear recoil and the capture signals,
are compared to data. In the relevant time interval 
used for the selection, the normalized predicted distributions 
describe the data quite well. 
This confirms that the selection procedure is reasonable.

\section{Results}
\label{sec:results}

The directly measured distributions related to the larger sample of
capture events do not require further analysis and are, therefore,
presented first.
They can be directly compared to MC predictions.
The nuclear recoil events were subject to further analysis which allowed the
extraction of the neutron spectrum. The comparison of MC predictions
to the result needs special consideration and is presented at the end.

\subsection{Capture events}
\label{sec:captured neutron}

\begin{figure}[h]
\centering
\includegraphics[width=0.48\textwidth]{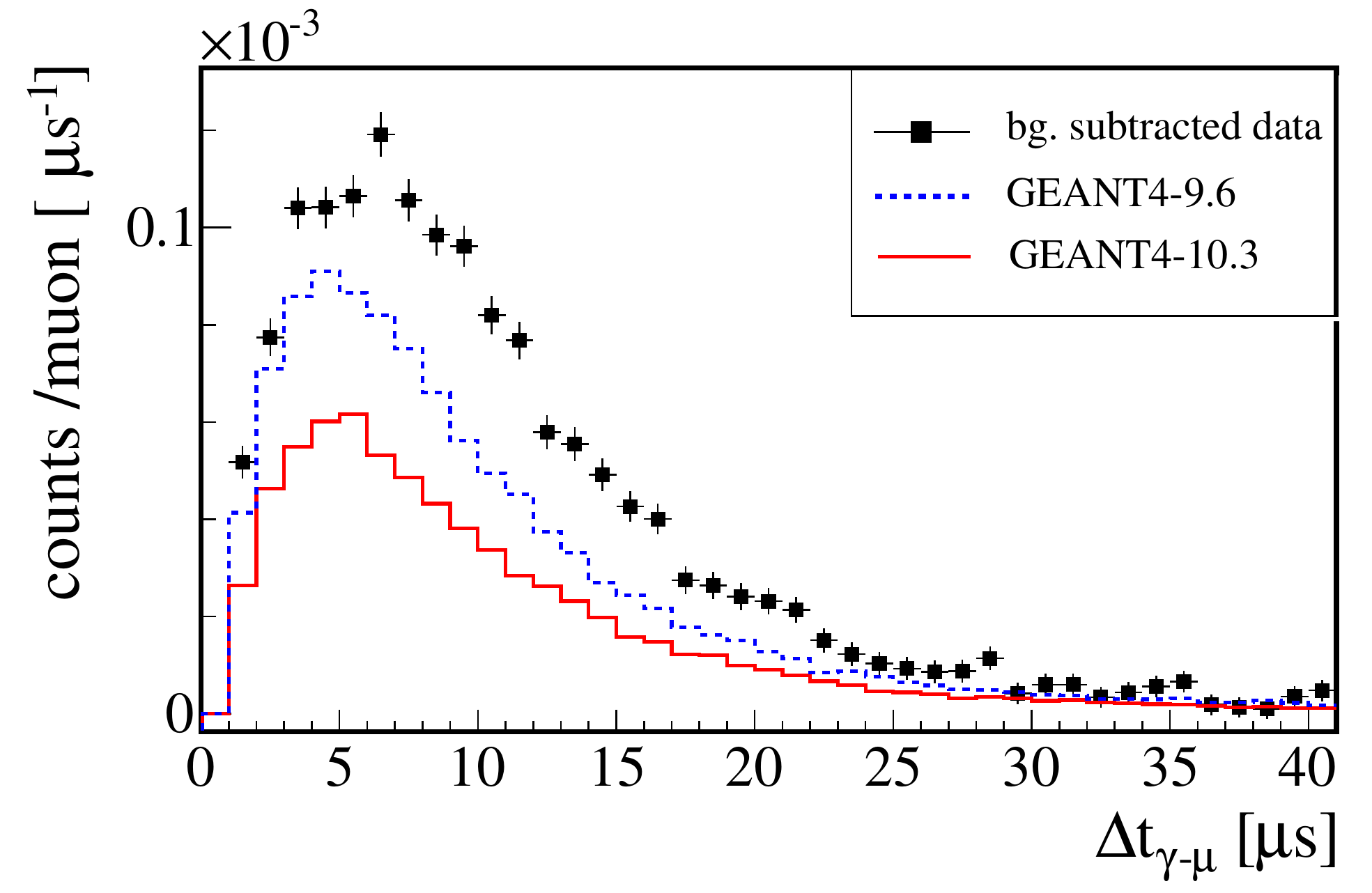}
\caption{\label{fig:capture-time} 
The distributions of the time delay,
$\Delta t_{\upgamma\textrm{-}\upmu}$,  
of the neutron capture signals with $E_{thr}^{\textrm{CE}}$\,=\,2\,MeV 
with respect to the muon tags as observed and as predicted by
GEANT4-9.6 and  GEANT4-10.3.
The distributions were normalized to the number of muon tags.}
\end{figure}

The time distribution of capture events was already 
shown in Fig.~\ref{fig:capture time}. After background subtraction,
MC predictions normalized to the number of muon tags 
can be directly compared to the data. The result is
shown in Fig.~\ref{fig:capture-time} for the predictions obtained 
with GEANT4-9.6 and GEANT4-10.3.
The time development of the signal is quite well described.
Contrary to this, the predicted overall signal strength is 
significantly too low for both MC versions.
The relevant numbers are summarized in Table~\ref{tab:neutron yield}.

The only significant source of systematic uncertainty is the
uncertainty on the energy scale of the Gd--LS detector of 
$\approx 10$\,\%.
% 1.8 MeV: 18850±177(stat.) after subtracting accidental background.
% 2.2 MeV: 15793±151(stat.) 
%     3057/2/34643 = 0.044
This translates to a systematic uncertainty of 4.5\,\% on the number of
events. After taking the systematic uncertainty into account, the factors by which the observation exceeds the predictions
are
$ 1.65~\pm\,0.02\,\textrm{(stat.)}~\pm\,0.07\,\textrm{(syst.)} $ 
and 
$ 2.58~\pm\,0.03\,\textrm{(stat.)}~\pm\,0.11\,\textrm{(syst.)} $ 
for GEANT4-9.6 and GEANT4-10.3, respectively.

The excess factors for increasing $E_{thr}^{\textrm{CE}}$ are listed in 
Table~\ref{tab:different factor}. The systematic uncertainty of 4.5\,\% 
affects the factors in the same way for all values of  $E_{thr}^{\textrm{CE}}$.
Thus, only statistical uncertainties are considered. 
The discrepancy between observation and predictions increases with
increasing energy threshold. This indicates that the spectrum of the
delayed $\upgamma$ must be harder than predicted.

\begin{table*}
\centering
\caption{\label{tab:neutron yield} Summary of observed and predicted
event rates for all neutron capture events with an experimental
energy threshold of $E_{thr}^{\textrm{CE}}$\,=\,2\,MeV and for
events with an identified nuclear recoil. The uncertainties for all 
muon capture events are statistical only.}
        \begin{tabular*}{\textwidth}{@{\extracolsep{\fill}}lccc}
		\toprule 	
    & experiment  & \tabincell{c}{GEANT4-9.6\\(Shielding)} &  \tabincell{l}{GEANT4-10.3\\(ShieldingM)}  \\
 \toprule
 muon tags
    & $7.30 \times 10^{6}$  & $28.72\times10^{6} $ & $28.76\times10^{6} $ \\  
\hline
 Neutron capture events
    & $26399\pm162$         & $41309\pm203$       & $26406\pm162$        \\
%  \hline
 Background
    & $9082\pm21$                   & --                    & --         \\
%			\hline                             	
 \tabincell{l}{Background subtracted events ($\times10^{3}$)}
    & $17.32\pm0.16$  & $41.31\pm0.20$    
    & $26.41\pm0.16$                                        \\
%			\hline
 \tabincell{l}{Neutron captures per muon tag ($\times10^{-3}$)}
    & $2.37\pm0.02$ & $1.438\pm0.007$ 
    & $0.918\pm0.006$                                     \\
%		    \hline
 Ratio observation/MC   & --      &  $1.65\pm0.02$ 
                         & $2.58\pm0.03$        \\	 
		\bottomrule

\hline
      
		    Nuclear Recoil Events  
		              &  $3534\pm68$                  & $6255\pm79$             & $3703\pm61$  \\
		    Events per muon tag ($\times10^{-4}$)
		              & $4.84\pm0.09$                 & $2.18\pm0.03$           & $1.29\pm0.02$\\
		    Systematic uncertainty  
		              &  11\%               & --                 & --           \\
		    Ratio observation/MC
		              & --   & \tabincell{l}{$2.22\pm0.05(\textrm{stat.})$\\\qquad$\pm\,0.25(\textrm{syst.})$}   
		                     & \tabincell{l}{$3.76\pm0.09(\textrm{stat.})$\\\qquad$\pm\,0.41(\textrm{syst.})$}                       
		   \\	
		   \bottomrule                             
  \end{tabular*}
\end{table*}

\begin{table}
\centering
\caption{\label{tab:different factor} 
Ratios of observed over predicted rates for
neutron capture events depending on the
experimental energy threshold $E_{thr}^{\textrm{CE}}$.
Uncertainties are statistical only.}
        \begin{tabular}{@{\extracolsep{\fill}}ccc}
		\toprule 		
  $E_{thr}^{\textrm{CE}}$   & GEANT4-9.6  &  GEANT4-10.3   \\
			\toprule
	0.5 MeV  & $1.71\pm0.02$       &    $2.60\pm0.04$         \\
%			\hline			
	1.0 MeV	 & $1.64\pm0.02$       &    $2.53\pm0.03$         \\
%			\hline
			1.5 MeV  			         & $1.64\pm0.02$       &    $2.57\pm0.03$         \\
%			\hline			
			2.0 MeV			             & $1.65\pm0.02$       &    $2.58\pm0.03$         \\
%			\hline
			2.5 MeV  			         & $1.67\pm0.02$       &    $2.58\pm0.03$         \\
%			\hline			
			3.0 MeV			             & $1.74\pm0.02$       &    $2.64\pm0.03$         \\
%			\hline
			3.5 MeV  			         & $1.87\pm0.03$       &    $2.71\pm0.04$         \\
%			\hline			
			4.0 MeV			             & $2.11\pm0.03$       &    $3.04\pm0.05$         \\
%			\hline
			4.5 MeV  			         & $2.34\pm0.04$       &    $3.23\pm0.06$         \\
		\bottomrule
        \end{tabular}
\end{table}

\begin{figure}
\centering
\includegraphics[width=0.48\textwidth]{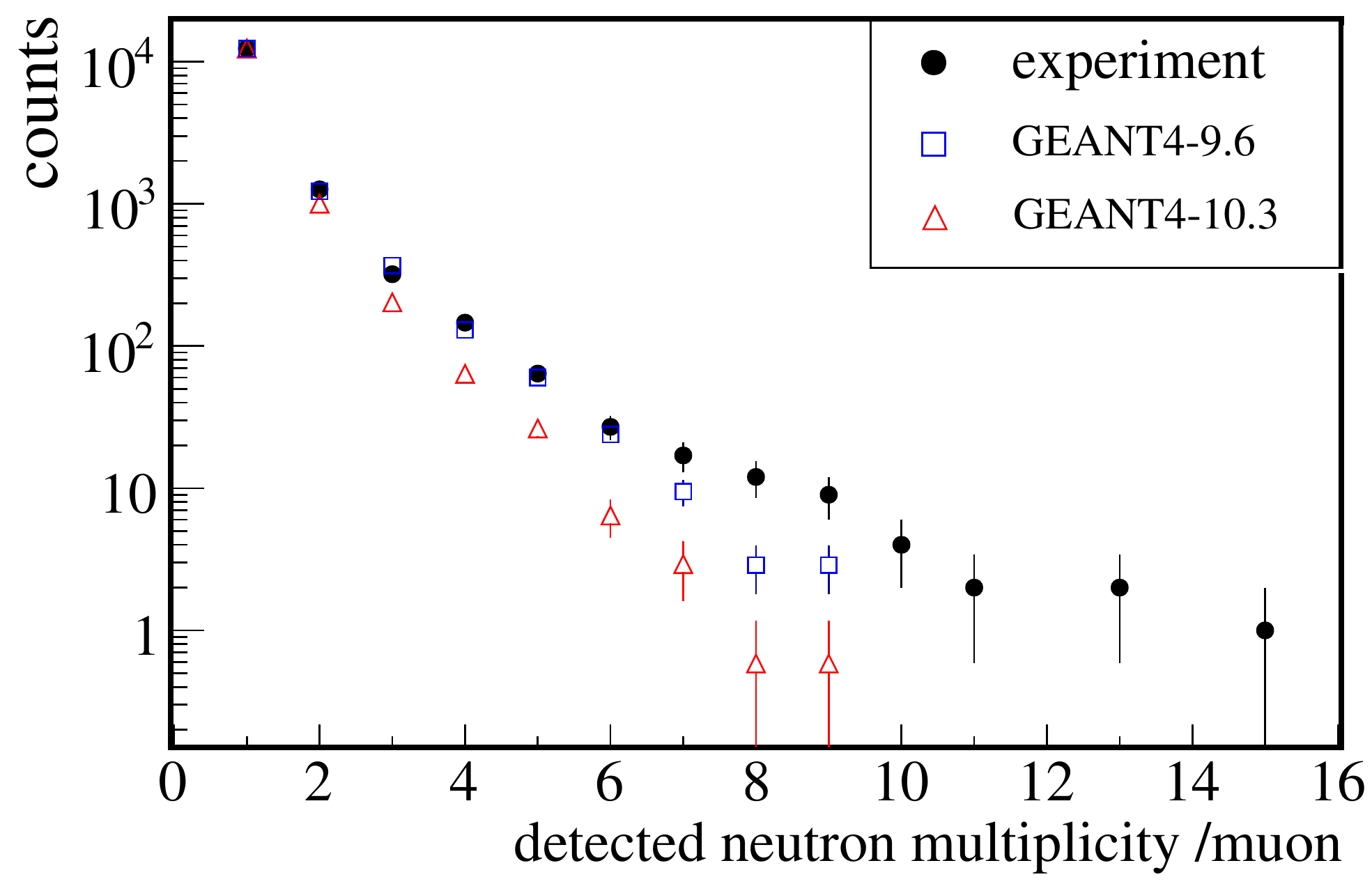}
\caption{\label{fig:multiplicity} The distribution 
of multiplicities for neutron capture signals with $E_{thr}^{\textrm{CE}}$\,=\,2\,MeV
as observed and predicted by GEANT4-9.6 and GEANT4-10.3.
The predictions were normalized by normalizing the 
predicted number of events with multiplicity one to the observed 
number of events with multiplicity one.  
%count of experiment except for the single multiplicity which is significantly influenced by the accidental background.
}
\end{figure}

The majority of the capture events, 71\,\%, have exactly
one signal indicating neutron capture. However, there are events with
more than one delayed $\upgamma$ in the 
time window 2\,$\upmu$s\,$<\Delta t_{\upgamma\textrm{-}\upmu}<$\,40\,$\upmu$s. 
This indicates that more than one neutron
was produced, reached the Gd--LS detector and was captured.
The distributions of detected multiplicities of neutron capture signals 
in events with one muon tag are shown in Fig.~\ref{fig:multiplicity}
as measured and as predicted by the two GEANT4 versions.  
The predictions were normalized by normalizing the predicted number of
events with multiplicity one to the background subtracted 
observed number of events with multiplicity one.
GEANT4-9.6 does better than GEANT4-10.3, which significantly underestimates
the number of high multiplicity events. 

\subsection{Nuclear recoil events}

\begin{figure}[t]
\centering
\includegraphics[width=0.48\textwidth]{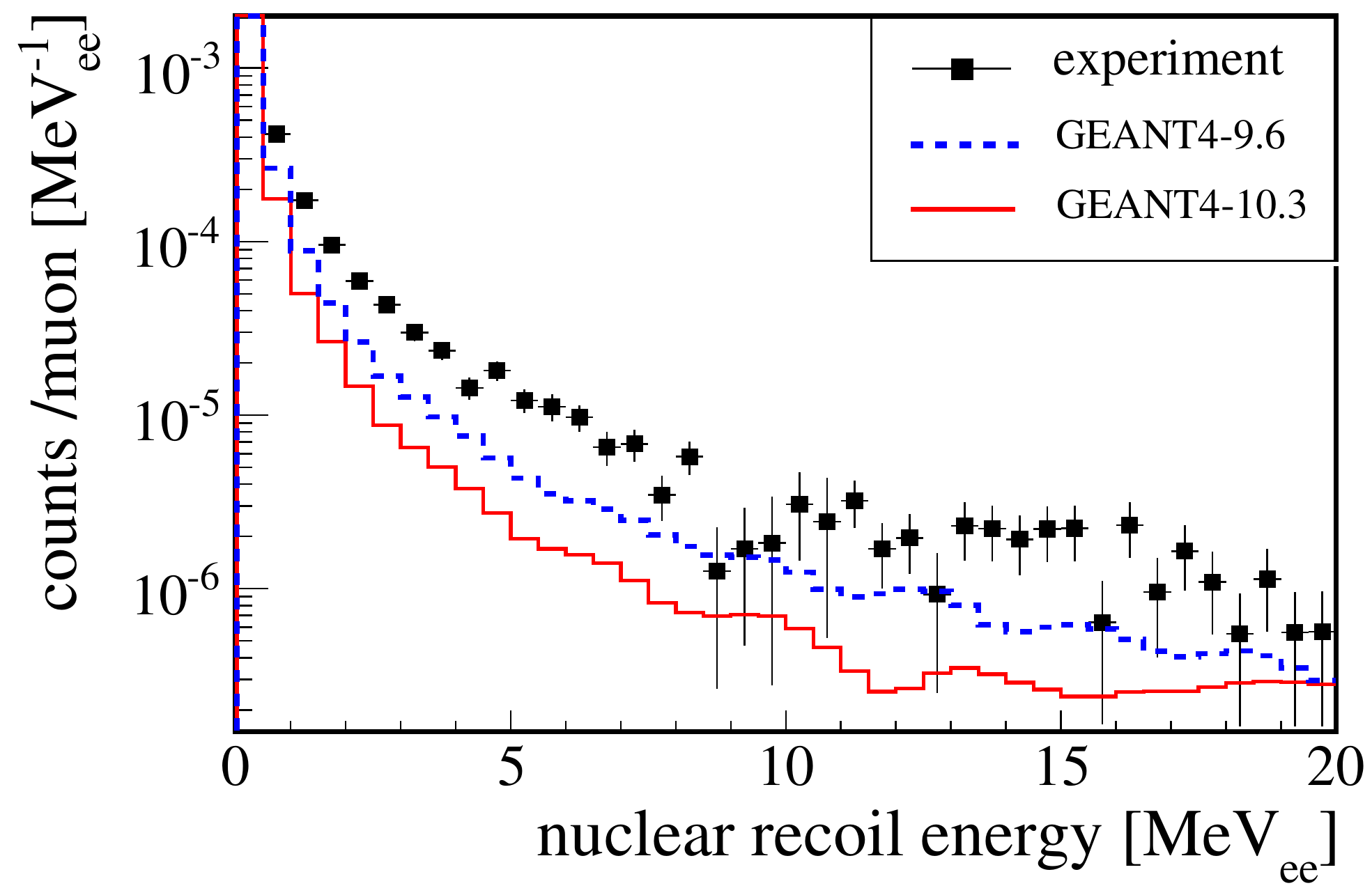}
\caption{\label{fig:spectrum of nuclear recoils} Spectra of nuclear
recoils as extracted for events with capture and nuclear recoil signals as
measured and predicted. The signal quenching for nuclear recoils in the Gd-LS
was part of the simulations. 
All spectra were normalized to the respective
number of muon tags. Only the statistical uncertainties of the
measurement are shown.
}
\end{figure}

The selection and treatment 
of events, in which in addition to the capture signal also
a signal indicating a prompt nuclear recoil was identified, 
%and the extraction of the signal per energy bin 
was described in detail in
Section~\ref{sec:data analysis}. 
The resulting nuclear recoil spectrum normalized to the
number of muon tags
is shown in Fig.~\ref{fig:spectrum of nuclear recoils}, obtained from the double-Gaussian fits described in Section~\ref{sec:final nuclear recoil}.
Also shown in Fig.~\ref{fig:spectrum of nuclear recoils} are the
predictions of the two MC versions.
As for all capture events, the predictions are significantly below
the number of observed nuclear recoils.
In the energy range of 0.5\,--\,20\,MeV$_{ee}$, 
the integrated number of predicted events 
is a factor of 
$2.22~\pm\,0.05\,\textrm{(stat.)}~\pm\,0.24\,\textrm{(syst.)}$ 
too low for GEANT4-9.6.
For GEANT4-10.3, the factor is 
$3.76~\pm\,0.09\,\textrm{(stat.)}~\pm0.41\,\textrm{(syst.)}$. 
The relevant numbers are listed in Table~\ref{tab:neutron yield}.

The deficit in the GEANT4 predictions for this subset of the 
neutron events is larger than for the complete neutron capture data set.
As the identification of a nuclear recoil requires a neutron energy of 
at least about 5\,MeV, this already suggests 
that the neutron spectrum is harder than predicted.

\subsection{Neutron spectrum}

Due to the basically background free determination of the nuclear
recoil spectrum, it was possible to actually measure the neutron spectrum
by unfolding the nuclear recoil spectrum.
To do so, the energy dependent response functions of the Gd-LS detector 
were determined by GEANT4 simulation 
as demonstrated in Fig.~\ref{fig:response function}
for mono-energetic neutrons
produced homogeneously inside the right-side wall. 
The simulation was performed for
neutron energies from 1 to 50\,MeV in 0.5\,MeV steps.

The simulated response functions were used as input to the iterative
unfolding method SAND-II~\cite{SAND}. 
A vector $\Phi_{i}^J$ with $i$ indicating the energy bin ($i=1,..,I$)
and width $\Delta E_{i}$
is the result of the procedure after $J$ iterations; 
$\Phi^{j}_{i}$ represents the result after $j$ iterations.
Each iteration is performed as:

\begin{equation}
\label{eq:SAND-II}
\begin{split}
\Phi^{j+1}_{i} = \Phi^{j}_{i} exp \left( {\frac{ \sum_{k=1}^{K} W^{j}_{ik} ln (U_{k}^{j}) }{ \sum_{k=1}^{K} W^{j}_{ik} } }\right)\\
\textrm{for}~~ j=0,1,\cdots,J\\
\end{split}
\end{equation}
with
\begin{equation*}
W_{ik}^{j} = \frac{R_{ki}\Phi_{i}^{j}}{\sum_{i=1}^{I}R_{ki}\Phi_{i}^{j}}\frac{N_{k}^{2}}{\sigma_{k}^{2}},\quad
U_{k}^{j} = \frac{N_{k}}{\sum_{i=1}^{I} R_{ki} \Phi_{i}^{j} \Delta E_{i}}
\end{equation*}
where
$K$ is the number of bins of the nuclear recoil spectrum;
$R_{ki}$ is the response function at the $k^{th}$ bin 
for the $i^{th}$ energy of mono-energetic neutrons;
$N_{k}$ is the measured number 
of nuclear recoils for the $k^{th}$ bin 
with the uncertainty $\sigma_{k}$. 
A flat spectrum $\Phi_{i}^{0}$ was chosen to start the iteration. 
The number of iterations $J$, after which the procedure ends, is
determined with convergence monitoring.

The method was previously confirmed through the unfolding of a spectrum
from an AmBe neutron source using a similar detector with the same dimensions 
and filled with identical Gd--LS~\cite{Qiang2017}.

\begin{figure}
\centering
\includegraphics[width=0.48\textwidth]{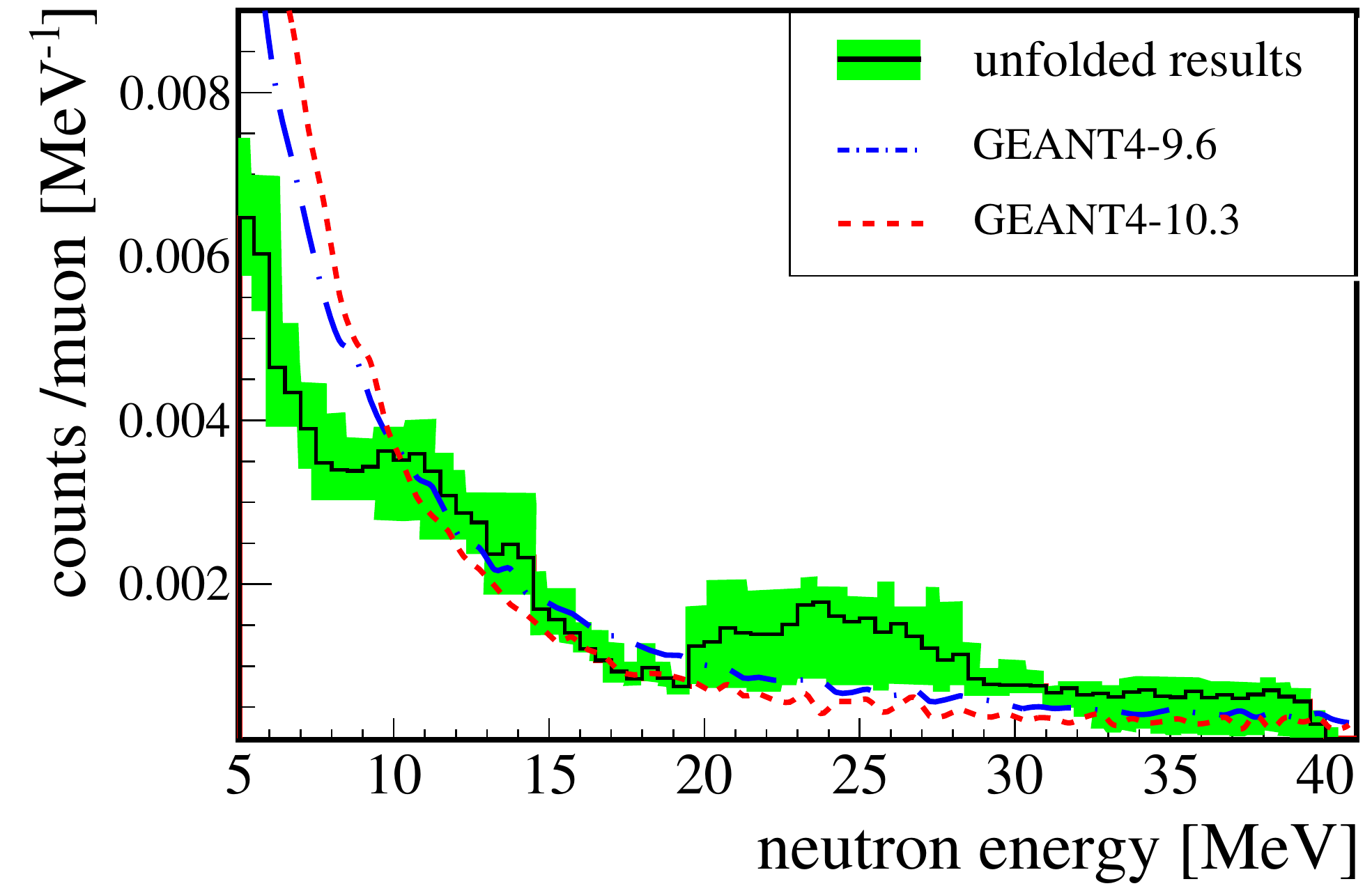}
\caption{\label{fig:unfolded spectrum} The unfolded spectrum of
neutrons induced by muons in lead (solid histogram). 
The band corresponds to the total  uncertainties, see text for details. 
Also shown are the predictions obtained with GEANT4-9.6 and GEANT4-10.3
normalized to the area of the unfolded spectrum from 5\,MeV to 40\,MeV.
}
\end{figure}

The result of the unfolding procedure is the neutron spectrum shown 
in Fig.~\ref{fig:unfolded spectrum}.
In order to estimate the combined statistical 
and systematic uncertainties, the unfolding 
was repeated with 
\begin{itemize}
\item  response functions obtained for 
       the simulated position and angular distributions 
       as shown in Fig.~\ref{fig:simulated distribution}; 
\item  a variation of the experimental energy scale of $\pm$\,10\,\%;

\item  The $N_k$ values varying by their statistical uncertainties.
\end{itemize}
The band in Fig.~\ref{fig:unfolded spectrum} represents the envelope of
all resulting changes to the spectrum.
Also shown in  Fig.~\ref{fig:unfolded spectrum}
are the predictions from the GEANT4 simulations
using the versions 9.6 and 10.3. 
The simulated spectra were normalized to the area of the 
measured spectrum in the energy range from 5\,MeV to 40\,MeV 
in order to compare the spectral shapes.
Even though the uncertainties are significant,
the observed spectrum is clearly harder than the predicted spectra.
This is more pronounced for GEANT4-10.3

\begin{figure*}
\centering
\begin{subfigure}{0.48\textwidth}
\includegraphics[width=\linewidth]{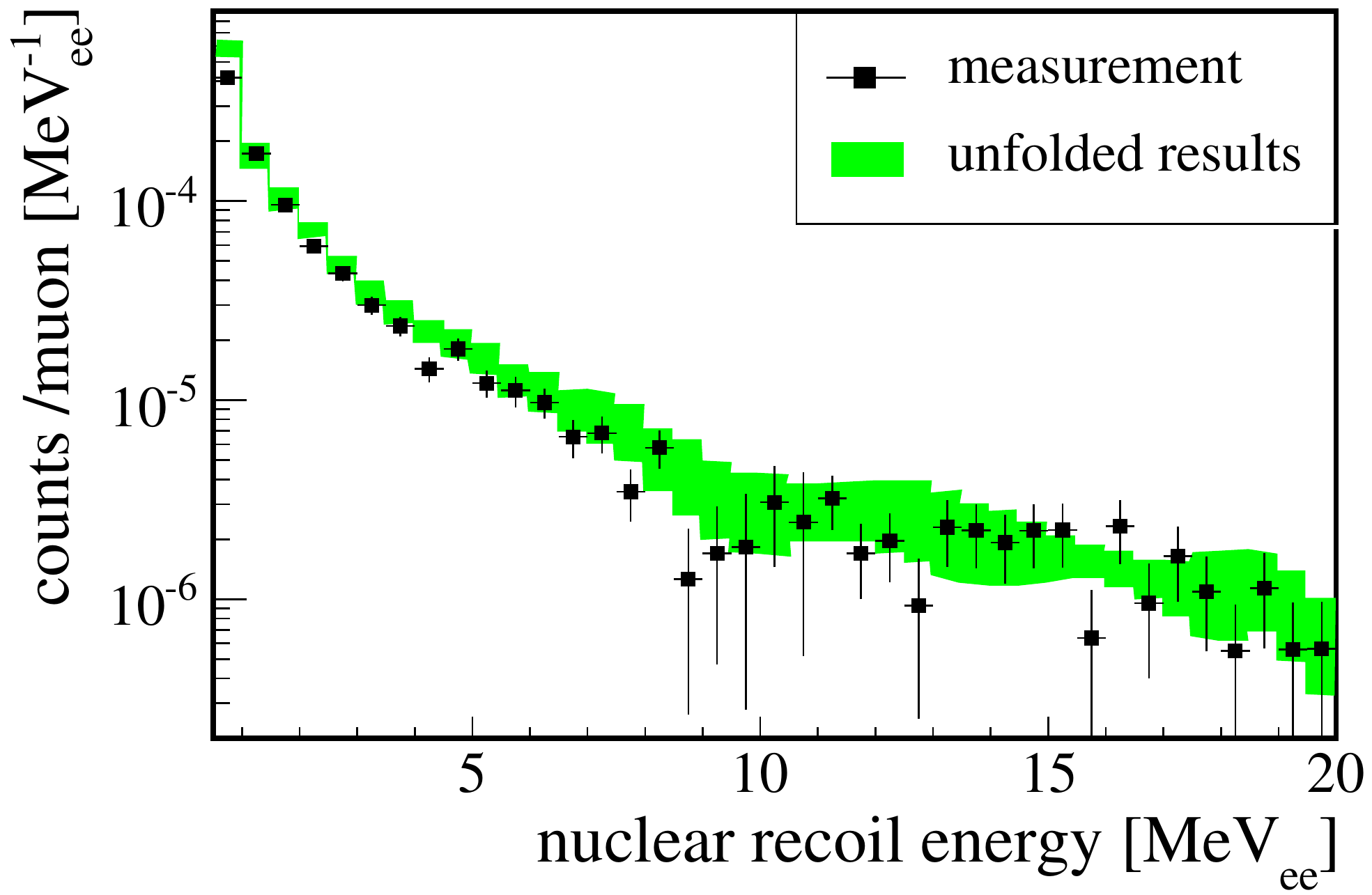}
\caption{\label{fig:recoil spectrum using unfold} }
\end{subfigure}
\hfill
\begin{subfigure}{0.48\textwidth}
\includegraphics[width=\linewidth]{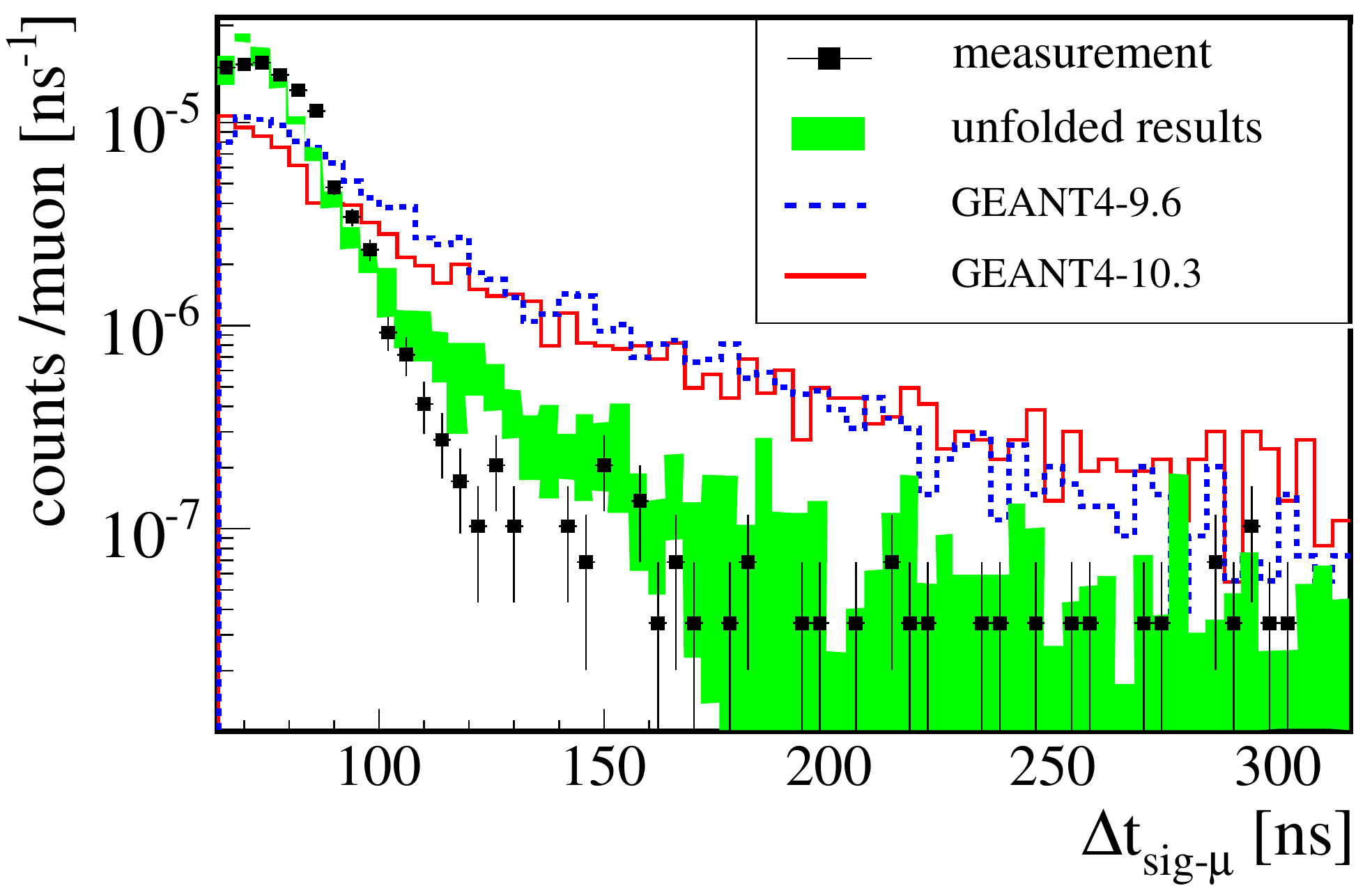}
\caption{\label{fig:time spectrum using unfold} }
\end{subfigure}
\caption{\label{fig:spectrum using unfold} (a) Nuclear recoil spectrum as
observed and as reproduced by GEANT4-9.6 from the unfolded neutron spectrum
as shown in Fig.~\ref{fig:unfolded spectrum}. 
(b) Distribution of the time delay between muon tag and neutron nuclear recoil signal
as observed, as reproduced by GEANT4-9.6 from the unfolded neutron spectrum
and as predicted by GEANT4-9.6 and GEANT4-10.3.
The bands represent the uncertainties due to the uncertainties on the
measured neutron spectrum used as input to the MC, see text for details.
}
\end{figure*}

As a consistency check, the unfolded neutron spectrum 
was used 
as the input to GEANT4-9.6 to obtain the nuclear recoil spectrum. 
The uncertainties as listed above were again considered by using the
respective neutron spectra as inputs.
The resulting band of predictions  is compared to the measurement 
in Fig.~\ref{fig:recoil spectrum using unfold}. 
The measured nuclear recoil spectrum is reproduced very well 
above 1\,MeV$_{ee}$.

The energy spectrum of the neutrons also influences the distribution of
the time difference $\Delta t_{\textrm{sig-}\upmu}$ between the muon tag
and the nuclear recoil signal.
The unfolded neutron spectrum and its uncertainties were again used as
input to GEANT4-9.6 to simulate the corresponding $\Delta t_{\textrm{sig-}\upmu}$
distribution. 
The result is shown in Fig.~\ref{fig:time spectrum using unfold} together
with the measured distribution and the distributions predicted by the two
GEANT4 versions. 
All three simulated time distributions were shifted 
by $+$48\,ns to account for the experimental time delay 
$\Delta t_{\textrm{delay}}^{\textrm{exp}}$, see Section~\ref{sec:nuclear recoil}.

The prediction resulting from the unfolded neutron spectrum describes
the data reasonably well. This demonstrates the consistency of the
unfolded neutron spectrum and the measured $\Delta t_{\textrm{sig-}\upmu}$
distribution.
It also confirms that the estimate of 
$\Delta t_{\textrm{delay}}^{\textrm{exp}}=48$\,ns is realistic.
The $\Delta t_{\textrm{sig-}\upmu}$ distributions
as predicted by the two  GEANT4 versions
do not describe the data. 
As GEANT4-9.6 describes the data quite well with the unfolded neutron
spectrum as input, the discrepancy can be attributed to the predicted
neutron spectrum and not to problems in the simulation of neutron transport.

\section{Discussion on GEANT4 performance}

The number of observed neutron captures and neutron nuclear recoils
is significantly higher than predicted by the two GEANT4 versions.
In addition, the predicted neutron spectrum is too soft.
Comparing the performance of the two GEANT4 versions it is observed
that not only the overall deficit of neutrons is even higher 
for the newer version GEANT4-10.3, but also the neutron spectrum 
is even softer, see Fig.~\ref{fig:unfolded spectrum}. 
The deficit in predicted high-multiplicity events is also larger
for GEANT4-10.3, see Fig.~\ref{fig:multiplicity}.

The overall deficits cannot be explained without a considerable
deficit in neutrons produced in muon-nucleon interactions.
In addition, the problems with
the angular distribution of muon-induced neutrons in GEANT4 
as observed by NA55~\cite{M. G. Marino} might contribute.

The deficit in the prediction of high multiplicity events, 
see Fig.~\ref{fig:multiplicity},
especially pronounced for GEANT4-10.3,  could be due to primary
production or due to a deficit in secondary neutron production.
Probably, the corresponding physics processes affect both.
The softness of the predicted neutron spectra could also cause 
a deficit in secondary particle production.   
The lack of predicted high-multiplicity events 
confirms the overall weaker performance of GEANT4-10.3. 

The softness of the predicted neutron spectra has to be linked to
the implementation of the physics processes relevant for muon-induced
neutron production. This is confirmed in that the time structure 
of nuclear recoil 
events can only be reproduced if the measured spectrum is used
in the Monte Carlo, see Fig.~\ref{fig:time spectrum using unfold}.   

The observation of a harder than predicted neutron spectrum is significant,
even though the measured nuclear recoil spectrum 
see Fig.~\ref{fig:spectrum of nuclear recoils},
from which it is unfolded,  
does not have high-energy information due
to the selection cut of 20\,MeV$_{ee}$. 
This limitation has to influence the high-energy part of  
the unfolded neutron spectrum somewhat. 
However, as the usage of the unfolded neutron spectrum in GEANT4 reproduces
the time structure of the observed nuclear recoil events, the influence
has to be small,
at least in the neutron energy range of 5\,--\,40\,MeV. 

The results presented here can hopefully help to improve the GEANT4 simulations.
The excess in observed rates cannot be used to
scale predictions for deep underground laboratories because the muon spectra
there are shifted towards significantly higher energies.
The measurements cannot be repeated at such deep depths
because of a lack of muon rate.
Thus, the physics process have to be better understood to make reliable 
predictions.

\section{Summary and conclusions}

Neutrons induced by cosmic muons in lead were studied in detail with
a Gadolinium doped liquid scintillator detector which was installed in the
shallow underground laboratory 
operated by the University of T\"ubingen,
next to the 
Muon-Induced Neutron Indirect Detection EXperiment. The MINIDEX plastic scintillators were used for muon tagging.
Samples of events with a neutron capture signal and with a prompt nuclear
recoil signal and a neutron capture signal were established with high
statistical significance.

Predictions made with the GEANT4 versions 9.6 and 10.3 were compared to
the data.
For neutron capture events, the observation exceeds the predictions by
factors of
$ 1.65\,\pm\,0.02\,\textrm{(stat.)}\,\pm\,0.07\,\textrm{(syst.)} $ 
and 
$ 2.58\,\pm\,0.03\,\textrm{(stat.)}\,\pm\,0.11\,\textrm{(syst.)} $ 
for GEANT4-9.6 and GEANT4-10.3, respectively.
For neutron nuclear recoil events, 
which require neutron energies above approximately
5\,MeV, the factors are even larger,
$ 2.22\,\pm\,0.05\,\textrm{(stat.)}\,\pm\,0.25\,\textrm{(syst.)} $ 
and 
$ 3.76\,\pm\,0.09\,\textrm{(stat.)}\,\pm\,0.41\,\textrm{(syst.)} $,
respectively.

The multiplicity of neutron captures per muon was investigated.
Events with multiplicity as high as 15 were observed.
Neither GEANT4 version predicts events with such high
multiplicities. Version 9.6 does quite well up to multiplicities
of 6 while version 10.3 already fails at a multiplicity of 3.

The energy spectrum of nuclear recoils was also observed with
high precision for recoil energies up to 20\,MeV$_{ee}$.
This recoil spectrum was used to obtain the spectrum of neutrons
in lead by unfolding. 
This represents the first statistically significant
measurement of the spectrum of neutrons
induced by cosmic muons in lead in a shallow underground laboratory.

The observed spectrum is harder than the spectra predicted by the
two GEANT4 versions. 
The time structure of the nuclear recoil events is not described
by the MC as is. However, if the observed spectrum
is used as an input, the time structure is reasonably well described.
This indicates that the problems with the GEANT4 description of the data arise
from problems with the implementation of the processes 
contributing to the creation of neutrons 
and not with neutron transport.
The hope is that the results presented here on the rate of neutron
production and on the neutron
spectrum can help to improve the GEANT4 simulations.

\section*{Acknowledgments}
This work was supported by the China Scholarship Council (CSC) (Contract No. 201606240105) and the Max-Plank-Society.



%%  \nocite{*}
%%  \bibliography{jos}

\section*{References}

 \begin{thebibliography}{00}

 \bibitem{GERDA} M. Agostini et al. (GERDA Collaboration), \href{https://www.nature.com/nature/journal/v544/n7648/full/nature21717.html}{Nature 544 (2017) 47-52}.
 
 \bibitem{PDG} C. Patrignani et al. (Particle Data Group), \href{http://iopscience.iop.org/article/10.1088/1674-1137/40/10/100001/meta}{Chin. Phys. C 40 (2016) 100001}. 

 \bibitem{dark matter} J. L. Liu et al., \href{https://www.nature.com/nphys/journal/v13/n3/abs/nphys4039.html}{Nat. Phys. 13 (2017) 212–216}.
 
 \bibitem{neutrino} Q. R. Ahmad et al. (SNO Collaboration), \href{https://journals.aps.org/prl/abstract/10.1103/PhysRevLett.87.071301}{Phys. Rev. Lett. 87 (2001) 071301}; S. Fukuda et al. (Super-Kamiokande Collaboration), \href{https://doi.org/10.1103/PhysRevLett.86.5656}{Phys. Rev. Lett. 86 (2001) 5656}; F. P. An et al. (Daya Bay Collaboration), \href{http://journals.aps.org/prl/abstract/10.1103/PhysRevLett.108.171803}{Phys. Rev. Lett. 108 (2012) 171803}.
 
 \bibitem{MAJORANA} W. Xu et al. (Majorana Collaboration), \href{http://iopscience.iop.org/article/10.1088/1742-6596/606/1/012004/meta#citations}{J. Phys. Conf. Ser. 606(1) (2015) 012004}.
 
 \bibitem{CDEX} K. J. Kang et al., \href{http://link.springer.com/article/10.1007/s11467-013-0349-1}{Front. Phys. 8 (2013) 412}; W. Zhao et al. (CDEX Collaboration), \href{http://journals.aps.org/prd/abstract/10.1103/PhysRevD.93.092003}{Phys. Rev. D 93 (2016) 092003}; S. K. Liu et al. (CDEX Collaboration), \href{https://journals.aps.org/prd/abstract/10.1103/PhysRevD.95.052006} {Phys. Rev. D 95 (2017) 052006}.
  
 \bibitem{1802.05040} C. Wiesinger, L. Pandola and S. Schönert, \href{https://arxiv.org/abs/1802.05040}{arXiv:1802.05040[hep-ex]}
 
 \bibitem{Boulby-1} H. M. Ara\'ujo et al., \href{https://doi.org/10.1016/j.astropartphys.2008.05.004}{Astroparticle Physics 29 (2008) 471–481}.
 
 \bibitem{H. M. Kluck} H. M. Kluck, Measurement of the cosmic-induced neutron yield at the Modane underground laboratory, PhD Thesis KIT, Karlsruhe (2013). \href{http://inspirehep.net/record/1351503?ln=en}{DOI: 10.1007/978-3-319-18527-9}.

 \bibitem{Gorshkov1974} G. V. Gorshkov, V. A. Zyabkin, and R. M. Yakovlev, Trans. by J. G. Adashko, Sov. J. Nucl. Phys 18.1 (1974) 57–61, Orig. pub. as Yad. Fiz. 18 (1973) 109–117 [in Russian]. 
 
 \bibitem{Iris2017} I. Abt et al., Astropart. Phys. 90, 1 (2017). \href{https://doi.org/10.1016/j.astropartphys.2017.01.011}{DOI: 10.1016/j.astropartphys.2017.01.011}  
 
 \bibitem{Crouch1952} M. F. Crouch and R. D. Sard, Phys. Rev., 2nd ser., 85.1 (1952) 120–129. \href{https://journals.aps.org/pr/pdf/10.1103/PhysRev.85.120}{DOI: 10.1103/PhysRev.85.120}. 
 
 \bibitem{Annis1954} M. Annis et al., Phys. Rev. 94, 4 (1954). \href{https://journals.aps.org/pr/pdf/10.1103/PhysRev.94.1038}{DOI: 10.1103/PhysRev.94.1038}.
  
 \bibitem{Gorshkov1971a} G. V. Gorshkov, V. A. Zyabkin, and R. Μ. Yakovlev, Trans. by C. S. Robinson, Sov. J. Nucl. Phys. 13.4 (1971) 450–452, Orig. pub. as Yad. Fiz. 13 (1971) 791–796 [in Russian].
 
 \bibitem{Holborn} A. M. Short, In Proceedings of the 9th International Cosmic Ray Conference, (London, UK), 1 (1965) 1009–1011. \href{http://articles.adsabs.harvard.edu//full/1965ICRC....2.1009S/0001011.000.html}{ADS: 1965IC RC....2.1009S}.  
 
 \bibitem{Bergamasco1970} L. Bergamasco, Nuovo Cim. B 66.1 (1970) 120–128. \href{https://link.springer.com/content/pdf/10.1007\%2FBF02710194.pdf}{DOI: 10.1007/BF02710194}. 

 \bibitem{Gorshkov1968} G. V. Gorshkov and V. A. Zyabkin, Trans. by J. G. Adashko, Sov. J. Nucl. Phys. 7.4 (1968) 470–474, Orig. pub. as Yad. Fiz. 7 (1968) 770–777 [in Russian].

 \bibitem{Gorshkov1971} G. V. Gorshkov and V. A. Zyabkin, Trans. by J. G. Adashko, Sov. J. Nucl. Phys. 12.2 (1971) 187–190, Orig. pub. as Yad. Fiz. 12 (1970), 340–346 [in Russian].

 \bibitem{Boulby-2} L. Reichhart et al., \href{https://doi.org/10.1016/j.astropartphys.2013.06.002}{Astroparticle Physics 47 (2013) 67–76}.

 \bibitem{Bergamasco1973} L. Bergamasco, S. Costa, and P. Picchi, Nuovo Cimento A 13 (1973) 403–412. \href{https://link.springer.com/content/pdf/10.1007\%2FBF02827344.pdf}{DOI: 10.1007/BF02827344}.
 
 \bibitem{E665} M. R. Adams et al. [E665 Collaboration], Phys. Rev. Lett. 74 (1995) 5198. \href{https://doi.org/10.1103/PhysRevLett.74.5198}{DOI: 10.1103/PhysRevLett.80.2020}.

 \bibitem{NA55} V. Chazal et al., Nucl. Instrum. Meth. A 490 (2002) 334–343. \href{https://doi.org/10.1016/S0168-9002(02)01006-9}{DOI: 10.1016/S0168-9002(02)01006-9}.

 \bibitem{M. G. Marino} M. G. Marino et al, Nucl. Instrum. Meth. A 582 (2007) 611–620. \href{https://doi.org/10.1016/j.nima.2007.08.170}{DOI: 10.1016/j.nima.2007.08.170}.

 \bibitem{Y. Nakajima} Y. Nakajima et al, AIP Conference Proceedings 1672, 090002 (2015). \href{http://dx.doi.org/10.1063/1.4928000}{DOI: 10.1063/1.4928000}.
  
  \bibitem{Depth} P. Grabmaier, Universit\"at T\"ubingen, private communication.

  \bibitem{EJ-335}\href{http://www.eljentechnology.com/products/liquid-scintillators/ej-331-ej-335}{www.eljentechnology.com/products/liquid-scintillators/ej-331-ej-335}.
  
  \bibitem{geant4}\href{http://geant4.cern.ch}{http://geant4.cern.ch}

 \bibitem{Boulby} E. Tziaferi et al., \href{http://www.sciencedirect.com/science/article/pii/S0927650506001873}{Astropart. Phys. 27 (2007) 326}.
 
  \bibitem{Qiang2017} Q. Du et al, \href{https://doi.org/10.1016/j.nima.2018.01.098}{Nucl. Instrum. Meth. A 889 (2018) 105–112}.
 
 \bibitem{fluka-1} T. T. Boehlen et al., \href{https://doi.org/10.1016/j.nds.2014.07.049}{Nuclear Data Sheets 120 (2014) 211-214}. 
 \bibitem{fluka-2} A. Fass\`o et al., \href{http://www.slac.stanford.edu/pubs/slacreports/reports16/slac-r-773.pdf}{CERN-2005-10, INFN/TC-05/11, SLAC-R-773 (2005)}.

 \bibitem{Bogdanova} L. N. Bogdanova et al., \href{https://doi.org/10.1134/S1063778806080047}{Phys. Atom. Nucl 69 (2006) 1293-1298}. \href{https://arxiv.org/abs/nucl-ex/0601019}{arXiv:nucl-ex/0601019}
 
 \bibitem{PDG-2014} K. A. Olive et al. (Particle Data Group), \href{http://dx.doi.org/10.1088/1674-1137/38/9/090001}{Chin. Phys. C 38 (2014) 090001}.
 
 \bibitem{Birks} J. B. Birks, \href{http://www.sciencedirect.com/science/book/9780080104720}{\emph{The Theory and Practice of Scintillation  Counting}, Pergamon, New York, (1964)}.
 
 \bibitem{Qiang-Quenching} Q. Du et al., \href{https://doi.org/10.1088/1748-0221/13/04/P04001}{JINST 13 (2018) P04001}. 

 \bibitem{Shielding1} GEANT4 Collaboration, \href{http://www.slac.stanford.edu/comp/physics/geant4/slac_physics_lists/shielding/shielding.html}{Shielding Physics List: for shielding, underground and high energy applications}; \href{http://www.slac.stanford.edu/comp/physics/geant4/slac_physics_lists/shielding/physlistdoc.html}{Shielding Physics List Description}
 
  \bibitem{BERT_HP} \href{http://geant4.cern.ch/support/proc_mod_catalog/physics_lists/referencePL.shtml}{GEANT4 Reference Physics Lists}

 \bibitem{Shielding2} GEANT4 Collaboration, \href{http://geant4.cern.ch/support/ReleaseNotes4.9.4.html}{GEANT4 9.4 Release Notes}; \href{http://geant4.cern.ch/support/ReleaseNotes4.9.5.html}{GEANT4 9.5 Release Notes}; \href{http://geant4.web.cern.ch/geant4/support/ReleaseNotes4.9.6.html}{GEANT4 9.6 Release Notes}. %\href{http://geant4.cern.ch/support/ReleaseNotes4.10.0.html}{Geant4 10.0 Release Notes};

 \bibitem{ShieldingM} M. Asai et al. (GEANT4 Collaboration), \href{https://geant4.web.cern.ch/geant4/results/papers/geant4_mc2015_final.pdf}{The GEANT4 Version 10 Series}.
 
 \bibitem{SAND} S. L. Wang et al., \href{http://iopscience.iop.org/article/10.1088/1674-1137/33/5/012/meta}{Chin. Phys. C 33 (2009) 378}; W. N. McElroy et al., U.S. Air Force Weapons Laboratory Report AFWL-TR-67-41 (1967); H. Sekimoto and N. Yamamuro, Nucl. Sci. Eng. 80 (1982) 101; J. T. Routti and J. V. Sandberg, Rad. Prot. Dosim. 10 (1985) 103.

 \end{thebibliography}


\end{document}